\documentclass[prd,twocolumn,reprint,preprintnumbers,nofootinbib]{revtex4-1}

\RequirePackage[colorlinks=true
,urlcolor=blue
,anchorcolor=blue
,citecolor=blue
,filecolor=blue
,linkcolor=blue
,menucolor=blue
,linktocpage=true
,pdfproducer=medialab
,pdfa=true
]{hyperref}

\usepackage{amsmath,amssymb,amsthm,amsfonts}
\usepackage{graphicx,tabularx}
\usepackage{color}
\usepackage{multirow}
\usepackage{comment}
\usepackage{enumitem}

\setlist[description]{leftmargin=0.4cm}
\setlist[itemize]{leftmargin=0.4cm}

\newcommand{\be}{\begin{equation} \begin{aligned}}
\newcommand{\ee}{\end{aligned} \end{equation}}
\newcommand{\beqa}{\begin{eqnarray}}
\newcommand{\eeqa}{\end{eqnarray}}

\newcommand{\ifb}{\text{fb}^{-1}}
\newcommand{\gev}{\text{GeV}}
\newcommand{\tev}{\text{TeV}}
\newcommand{\mev}{\text{MeV}}
\newcommand{\sba}{s_{\beta-\alpha}}
\newcommand{\cba}{c_{\beta-\alpha}}

\newcommand{\tb}{t_{\beta}}
\newcommand{\tanb}{\tan \beta}

\def\figureautorefname~#1\null{Fig.\,#1\null}
\def\tableautorefname~#1\null{Tab.\,#1\null}

\def\equationautorefname~#1\null{Eq.\,(#1)\null}

\RequirePackage[normalem]{ulem}

\begin{document}
\title{2HDM Neutral Scalars under the LHC}

\author{Felix Kling}
\email{felixk@slac.stanford.edu}
\affiliation{SLAC National Accelerator Laboratory, 2575 Sand Hill Road, Menlo Park, CA 94025, USA}

\author{Shufang Su}
\email{shufang@email.arizona.edu}
\affiliation{Department of Physics, University of Arizona, Tucson, AZ  85721, USA}

\author{Wei Su}
\email{wei.su@adelaide.edu.au}
\affiliation{ARC Centre of Excellence for Particle Physics at the Terascale, Department of Physics,University of Adelaide, South Australia 5005, Australia}

\begin{abstract}
Two Higgs Doublet Models (2HDM) provide a simple framework for new physics models with an extended Higgs sector. The current LHC results, including both direct searches for additional non-Standard Model (SM) Higgs bosons, as well as precision measurements of the SM-like Higgs couplings, already provide strong constraints on the 2HDM parameter spaces.  In this paper, we examine those constraints for the neutral scalars in the Type-I and Type-II 2HDM.  In  addition to the direct search channels with SM final states: $H/A \to f\bar f, VV, Vh, hh$, we study in particular the exotic decay channels of $H/A \to AZ/HZ$ once there is a mass hierarchy between the non-SM Higgses.  We found that $H/A \to AZ/HZ$ channel has unique sensitivity to the alignment limit region which remains unconstrained by conventional searches and Higgs precision measurements. This mode also extends the reach at intermediate $\tb$ for heavy $m_A$ that are not covered by the other direct searches. 
\end{abstract}

\maketitle
\renewcommand{\baselinestretch}{0.9}\normalsize
\tableofcontents
\renewcommand{\baselinestretch}{1.0}\normalsize

\section{Introduction}

Since the discovery of the Standard Model (SM)-like 125 GeV Higgs boson at the first run of  the Large Hadron Collider (LHC)~\cite{Aad:2012tfa,Chatrchyan:2012xdj}, the SM is confirmed to be a self-consistent theory. Meanwhile motivated by various experimental observations and theoretical considerations, such as the existence of dark matter, the baryon asymmetry of the universe, the strong CP problem, the muon $g$-2 anomaly, or the naturalness problem, searches for new physics beyond the SM still remain a frontier in particle physics research. 

Many of the proposed new physics models contain an extended Higgs sector, among which the Two Higgs Doublet Model (2HDM) is one of the simplest options.  After electroweak symmetry breaking (EWSB), the general CP-conserving 2HDM contains five physical eigenstates: the observed 125 GeV CP-even neutral scalar $h$, an additional CP-even neutral scalar $H$, one CP-odd Higgs boson $A$, and a pair of charged Higgs boson $H^\pm$~\cite{Branco:2011iw}. The discovery of any beyond the SM (BSM) Higgses will be an unambiguous evidence for the existence of an extended Higgs sector.

There are usually two ways to probe extended Higgs sectors: through their modifications to the SM-like Higgs couplings  tested by  Higgs coupling precision measurements ~\cite{ATLAS:2019slw}, and direct searches  for BSM Higgses at high energy colliders \cite{Heinemeyer:2013tqa}. The current direct searches at the LHC include  both conventional search channels of  $H/H^\pm/A \to f\bar f$, $VV'$, as well as final states involving a SM-like Higgs $H \to Vh$, $hh$.  Under the alignment limit of the 2HDM, in which the 125 GeV Higgs $h$ is exactly SM-like, both  the decays $H\to V V$ and $H \to Vh$, $hh$  as well as the $VH$ and WBF production modes  vanish at tree level, making a discovery more challenging. However, if there is a mass hierarchy between the BSM Higgses, additional exotic decay modes, such as $H/A \to AZ/HZ$, $H/A\to H^\pm W^\mp$,  $H \to AA, H^+H^-$ or $H^\pm \to HW/AW$ open up and quickly dominate the decay branching fractions.  Such exotic decay modes open a new window to search for heavy BSM Higgses.  Meanwhile current collider limits on heavy Higgses would be relaxed given the suppression of their decay branching fractions.
In this paper we comprehensively examine the current constraints on the 2HDM parameter space in mass-degenerate and hierarchical scenarios, highlighting the complementarity and importance of the exotic Higgs decay channel $H/A \to HZ/AZ$.

The  rest  of  the  paper  is  organized  as  follows. We will give a brief introduction to 2HDM at \autoref{sec:2hdm} and compare the Type-I and Type-II. In \autoref{sec:conven}, we summarize the latest LHC searches that are relevant for Higgs studies . We show our interpreted results for the Type-I and Type-II 2HDM under the degenerate mass assumption in \autoref{sec:degenerate}. In \autoref{sec:exotic}  we extend this discussion to 2HDMs with non-degenerate mass spectra, focusing on the exotic decay channel of $H/A \to AZ/HZ$.  We conclude in \autoref{sec:con}.
 
\section{The 2HDM}
\label{sec:2hdm}

The scalar sector of the 2HDM consists of two $SU$(2) doublets $\Phi_i$, $i=1,2$, which can be parameterized as
\be
\Phi_i =
\left( \begin{array}{c}
\phi_i^+  \\
(v_i+\phi_i^0+i\varphi_i)/\sqrt{2}
\end{array} \right).
\ee
Here $v_i$ are the vacuum expectation values for the neutral components which satisfy the condition $v_1^2 + v_2^2 = v^2$ with $v=246~\gev$. After imposing a discrete $\mathcal{Z}_2$ symmetry on the Lagrangian to avoid tree-level flavour changing neutral currents (FCNCs), the 2HDM parameter space is described by six free parameters. For our purposes it is convenient to parametrize the 2HDM by the physical Higgs masses ($m_h$, $m_H$, $m_A$ and $m_{H^\pm}$), the mixing angle between the two CP-even Higgses ($\alpha$), and the ratio of the two vacuum expectation values ($\tb = v_2/v_1$). If we allow for a soft breaking of the $\mathcal{Z}_2$ symmetry, there is an additional parameter  $m_{12}^2$.

After EWSB, the scalar sector consists of five states: a pair of neutral CP-even Higgses, $h$ and $H$, a CP-odd Higgs, $A$, and a pair of charged Higgses $H^\pm$. For the neutral states we can write
\be
\centering
h &= - s_\alpha \phi_1^0 + c_\alpha \phi_2^0, \quad\\
H &= \phantom{-} c_\alpha \phi_1^0 + s_\alpha \phi_2^0,   \\
A &= - s_\beta \varphi_1 + c_\beta \varphi_2,  \quad
\ee
where we used the shorthand notation $s_x=\sin x$ and $c_x=\cos x$. In the following we will identify $h$ with the discovered $125~\gev$ Higgs\footnote{Note that here we use a convention in which $h$ is always the 125 GeV SM-like Higgs and the alignment limit is always at $\cba=0$. This is different from the mass ordered convention in which $h$ is the light CP-even Higgs and $H$ is the heavy one. }.   Note that in the generic 2HDM, Higgs masses are free parameters and therefore exotic Higgs decays such as $A \to HZ$ are possible when kinetically accessible.

In the context of this paper, we are mainly interested in the couplings of the neutral Higgses to the SM gauge bosons $V = Z, W^\pm$. The couplings of the neutral CP-even Higgses to a pair of vector bosons are
\be
g_{hVV}=\frac{m_V^2}{v} \sba \quad\text{and}\quad g_{HVV}=\frac{m_V^2}{v} \cba,
\ee
while the CP-odd Higgs $A$ does not couple to vector boson pairs. Additionally, the neutral CP-even Higgses can couple to the CP-odd Higgs $A$ and a $Z$-boson with couplings
\be
g_{hAZ}&= \phantom{-}\frac{g \cba }{2 c_{\theta_W}} (p_h-p_A)_\mu \quad\text{and} \quad \\
g_{HAZ}&= - \frac{g \sba }{2 c_{\theta_W}} (p_H-p_A)_\mu \ ,
\ee
where $\theta_W$ is the Weinberg angle and $p_\mu$ are the incoming Higgs momenta. LHC Higgs coupling measurements favor the \textit{alignment limit}, $\sba \approx 1$, in which the couplings of the $125~\gev$ Higgs to fermions and gauge bosons are consistent with those predicted by the SM \cite{Branco:2011iw,Chen:2018shg,Gu:2017ckc,Chen:2019pkq,Chen:2019rdk}.
In this case, the coupling of the BSM CP-even neutral Higgs $H$ to vector boson pairs is suppressed by $\cba \approx 0$, which is consistent with the non-observation of such a state in the $H\to VV$ channel \cite{Chen:2019bay,Su:2019dsf}. On the other hand, in the alignment limit, the CP-odd Higgs will couple more strongly to the BSM Higgs $H$ than its SM-like counterpart $h$. In particular, this implies that $A$ is more likely to decay to $HZ$ than $hZ$, if kinematically  possible. This motivates the exotic Higgs searches for $A \to HZ$ and $H \to AZ$ as complementary probe in the alignment limit.

Unlike the couplings to fermions and vector bosons, the triple and quartic Higgs couplings depend on the otherwise unobservable soft $\mathcal{Z}_2$ breaking parameter $m_{12}^2$. The corresponding expressions for various triple Higgs couplings have been obtained in~\cite{Kling:2016yls}. However, it has been shown in Ref.~\cite{Kling:2016opi} that satisfying unitarity and vacuum stability bounds for arbitrary values of $\tb$ requires $m^2_{12}=m^2_H s_\beta c_\beta$, with deviations possible only at $\tb\sim 1$. For illustration, we consider that this relation holds, in which case we can write the triple Higgs couplings as 
\be
\!\!g_{Hhh}\!=&\!-\!\frac{\cba}{v} \frac{s_{2\alpha}}{s_{2\beta}} (m_H^2\!-\! m_h^2)\!+\!\frac{\cba}{2v} m_H^2, \\    
\!\!g_{HAA}\!=&\!-\!\frac{\cba}{2v} (m_H^2\!-\!2 m_A^2),  \\
\!\!g_{hhh}\!=&\!-\!\frac{ \cba^2}{v}\! \left[\!\frac{\cba}{t_{2\beta}}\!-\!\sba\!\right]\!(m_H^2\!-\!m_h^2)\!+\! \frac{\sba }{2v}  m_h^2, \\
\!\!g_{hAA}\!=&\!-\!\frac{\sba}{2v} [2(m_H^2\!-\!m_A^2)\!-\!m_h^2] 
\!-\! \frac{\cba}{t_{2\beta} v} (m_H^2\!-\!m_h^2).\!\!\! 
\ee
We can see that in the alignment limit $\cba=0$, the decays of the heavy neutral Higgs $H\to hh, AA$ are absent and the SM Higgs self coupling obtains its SM value $g_{hhh}=m_h^2/(2v)$, while a decay of $h \to AA$ is possible if kinematically open. 

\begin{table}[t!]
\centering
\begin{tabular}{c|c|c|c|c}
  \hline
  \hline
  2HDM 		& Type-I 			& Type-II  			& Type-L 			& Type-F \\
  \hline
  \hline
  up-type		& $\Phi_2$			& $\Phi_2$			& $\Phi_2$			& $\Phi_2$			\\
  \hline
  $\xi_{huu}$	& $c_\alpha/s_\beta$ 	&$c_\alpha/s_\beta$ 	&$c_\alpha/s_\beta$	&$c_\alpha/s_\beta$ 	\\
  $\xi_{Huu}$ 	& $s_\alpha/s_\beta$ 	&$s_\alpha/s_\beta$	&$s_\alpha/s_\beta$ 	&$s_\alpha/s_\beta$ 	\\
  $\xi_{Auu}$	& $t_\beta^{-1}$ 	&$t_\beta^{-1}$  	&$t_\beta^{-1}$ 		&$t_\beta^{-1}$ 		\\
  \hline
  \hline
  down-type	& $\Phi_2$			& $\Phi_1$			& $\Phi_2$			& $\Phi_1$			\\
  \hline
  $\xi_{hdd}$	& $c_\alpha/s_\beta$ 	&$-s_\alpha/c_\beta$ 	&$c_\alpha/s_\beta$	&$-s_\alpha/c_\beta$ 	\\
  $\xi_{Hdd}$ 	& $s_\alpha/s_\beta$ 	&$c_\alpha/c_\beta$	&$s_\alpha/s_\beta$ 	&$c_\alpha/c_\beta$ 	\\
  $\xi_{Add}$	& $-t_\beta^{-1}$ 	&$t_\beta$  		&$-t_\beta^{-1}$ 	&$t_\beta$ 		\\
  \hline
  \hline
  lepton		& $\Phi_2$			& $\Phi_1$			& $\Phi_1$			& $\Phi_2$			\\
  \hline
  $\xi_{h\ell\ell}$	& $c_\alpha/s_\beta$ 	&$-s_\alpha/c_\beta$ 	&$-s_\alpha/c_\beta$	&$c_\alpha/s_\beta$ 	\\	
  $\xi_{H\ell\ell}$ & $s_\alpha/s_\beta$ 	&$c_\alpha/c_\beta$	&$c_\alpha/c_\beta$ 	&$s_\alpha/s_\beta$ 	\\	
  $\xi_{A\ell\ell}$	& $-t_\beta^{-1}$ 	&$t_\beta$  		&$t_\beta$ 		&$-t_\beta^{-1}$  	\\	
  \hline
  \hline
\end{tabular}
\caption{Types of 2HDM and the tree-level couplings of $h$, $H$ and $A$  to fermions, normalized to their SM Yukawa couplings: $\xi = y/y^{SM}$. }
\label{tab:2hdm-types}
\end{table}

As mentioned above, we have introduced a soft breaking $\mathcal{Z}_2$ symmetry to avoid tree-level FCNCs, which implies that each fermion type is only allowed to couple to one Higgs doublet. There are four possible types of 2HDMs: Type-I, Type-II, Type-L (or lepton-specific) and Type-F (or flipped), which we show in \autoref{tab:2hdm-types}. The corresponding couplings of the neutral scalar states to SM fermions normalized to their SM values, can be expressed in terms of the mixing angles $\alpha$ and $\beta$ and are also shown in~\autoref{tab:2hdm-types}. For the remainder of this paper we focus on 2HDMs of Type-I and Type-II, and the main differences of the interpreted results come from the BSM Higgs couplings to fermion pairs.

\section{Collider Constraints on BSM Higgses}
\label{sec:conven}

A variety of LHC measurements can be used to constrain extended Higgs sectors such as 2HDMs. This includes indirect constraints from precision Higgs couplings, direct searches for additional Higgses as well as  measurements of SM processes. In the following we summarize the different searches and measurements that we use for our analyses. Here we mostly focus on the neutral scalars, and comment on the charged scalar at the end.

\begin{description}
\item[Precision Higgs Measurements] 
    While the couplings of the 125 GeV Higgs $h$ are fixed in the SM, they are modified in the 2HDM: at tree level they depend on the mixing angles $\cba$ and $\tb$,  while additional dependence on the Higgs masses is induced via loop effects~\cite{Coleppa:2013dya, Gu:2017ckc, Chen:2018shg, Chen:2019rdk, Chen:2019bay,Chen:2019pkq}. We use the latest combined LHC 13 TeV measurements of the Higgs coupling strength at both CMS with $36~\ifb$~\cite{CMS:2018lkl} and ATLAS with $80~\ifb$~\cite{ATLAS:2019slw}. 
    
    Additionally, the Higgs width has been measured with high precision: $0.08~\mev<\Gamma_h<9.16~\mev$ at 95\% C.L., by CMS~\cite{Sirunyan:2019twz}. This measurement put strong constrains on both the enhanced couplings of $h$ to fermions at high/low $\tb$ and additional decay modes such as $h \to AA$. 
\item[Conventional channels] 
    Most of the existing direct searches for BSM Higgs bosons focus on their conventional decays into a pair of quarks, leptons or gauge bosons. The following table presents a summary of recent searches performed at the 13 TeV LHC. 
    %
    \begin{equation}
    \nonumber
    \begin{tabular}{c|c|c}
      \hline
      \hline
      channel 		        & CMS 			& ATLAS 	\\
      \hline
      $A/H \to \mu\mu$      
      & \cite{Sirunyan:2019tkw} 
      & \cite{Aaboud:2019sgt}
      \\
      $A/H \to bb$          
      & \cite{Sirunyan:2018taj}        
      & \cite{Aad:2019zwb}
      \\
      $A/H \to \tau\tau$        
      & \cite{Sirunyan:2018zut,CMS:2019hvr} 
      & \cite{Aad:2020zxo}
      \\
      $A/H \to \gamma\gamma$ 
      & \cite{Sirunyan:2018aui, Sirunyan:2018wnk}
      & \cite{Aad:2014ioa,Aaboud:2017yyg, ATLAS:2018xad}
      \\
      $A/H \to tt$ 
      & \cite{Sirunyan:2019wph} 
      & -
      \\
      $H \to ZZ$ 
      & \cite{Sirunyan:2018qlb}
      & \cite{Aaboud:2017rel}
      \\
      $H\to WW$ 
      & \cite{Sirunyan:2019pqw} 
      & \cite{Aaboud:2017gsl}\\
      \hline
      \hline
    \end{tabular}
    \end{equation}
    Note that the $gg\to A/H \to tt$ channel involves a non-trivial interference with the SM backgrounds that has been taken into account in the analysis. A brief summary of Higgs search results can be found in Ref.~\cite{CMS:2016qbe} for CMS at 8~TeV and Ref.~\cite{ATL-PHYS-PUB-2019-034} for ATLAS at 13~TeV.
\item[Exotic Decays into the SM Higgs] 
    Away from the alignment limit $\cba=0$, the heavy CP-odd Higgs can decay into the SM-like Higgs via $A \to hZ$. Additionally, the heavy CP-even Higgs can decay into a pair of SM-like Higgses, $H \to hh$. In our analysis we take into account the search results listed in the following table, where for the resonant di-Higgs channel,  we consider the limits resulting from the combination of different final states. 
    %
    \begin{equation}
    \nonumber
    \hspace{0.4cm}
    \begin{tabular}{c|c|c|c|c}
      \hline
      \hline
      channel& \multicolumn{2}{c|}{ATLAS} & \multicolumn{2}{c}{CMS} 
      \\
      \cline{2-5}
      & 8 TeV & 13 TeV
      & 8 TeV & 13 TeV 	\\
      \hline
      $A \to hZ \to bb\ell\ell$      
      & \cite{Khachatryan:2015lba}      
      & \cite{Sirunyan:2019xls}         
      & \cite{Aad:2015wra}              
      & \cite{Aaboud:2017cxo}           
      \\
      $A \to hZ \to \tau\tau\ell\ell$      
      & \cite{Khachatryan:2015tha}      
      & \cite{Sirunyan:2019xjg}         
      & \cite{Aad:2015wra}              
      & ---
      \\
      \hline
       $H \to hh $      
      & \cite{Sirunyan:2017tqo}         
      & \cite{Sirunyan:2018two}         
      & \cite{Aad:2015xja}              
      & \cite{Aad:2019uzh}              
      \\     
      \hline
      \hline
    \end{tabular}
    \end{equation}
%
\item[Exotic Decays of the SM Higgs] 
    If the BSM scalars are sufficiently light, $m_{A/H}<m_h/2$, exotic decays of the SM-like Higgs $h \to AA/HH$ open up. While the decay $h\to HH$ vanishes for $\cba=0$, the decay $h \to AA$ is unsuppressed under the alignment limit. The following table lists the current LHC searches considered for such channels, focusing on masses of  $m_A>4~\gev$. Searches listed in parenthesis do not provide the leading constraints and are only listed for completeness.
    %
    \begin{equation}
    \nonumber
    \hspace{0.4cm}
    \begin{tabular}{c|c|c}
      \hline
      \hline
      channel& ATLAS & CMS \\
      \hline
      $h\to AA \to bbbb$      
      & \cite{Aaboud:2018iil}      
      & ---                        
      \\
      $h\to AA \to bb\tau\tau$      
      & ---                        
      & \cite{Sirunyan:2018pzn}    
      \\
      $h\to AA \to bb\mu\mu$      
      & \cite{Aaboud:2018esj}      
      & \cite{Sirunyan:2018mot}    
      \\
      $h\to AA \to \tau\tau\tau\tau$      
      & ---             
      & \cite{Sirunyan:2019gou}    
      \\
      $h\to AA \to \tau\tau\mu\mu$      
      & \cite{Aad:2015oqa}         
      & \cite{Sirunyan:2018mbx}    
      \\
      $h\to AA \to \mu\mu\mu\mu$      
      & (\cite{Aaboud:2018fvk})    
      & (\cite{Sirunyan:2018mgs})  
      \\
      \hline
      \hline
    \end{tabular}
    \end{equation}
    %
    A brief summary on search results for exotic decays of the SM-like Higgs can be found in Ref.~\cite{Khachatryan:2017mnf} for CMS at 8 TeV and Ref.~\cite{ATL-PHYS-PUB-2018-045} for ATLAS at 13 TeV.
\item[Exotic Decays in BSM Sector] 
    If the two neutral BSM Higgs states $A$ and $H$ have a sufficient mass splitting, $|m_A-m_H|>m_Z$, the exotic decay channel $A/H \to HZ/AZ$ opens up. Both ATLAS and CMS have performed searches listed in the following table.
    %
    \begin{equation}
    \nonumber
    \hspace{0.4cm}
    \begin{tabular}{c|c|c}
      \hline
      \hline
      channel & ATLAS & CMS \\
      \hline
      $A/H\!\to\! HZ\!/\!AZ \!\to\! bb\ell\ell$      
      & \cite{Aaboud:2018eoy} ($13~\tev$)    
      & \cite{Sirunyan:2019wrn} ($13~\tev$)    
      \\
      $A/H\!\to\! HZ\!/\!AZ \!\to\! \tau\tau\ell\ell$      
      & ---                                              
      & \cite{Khachatryan:2016are} ($8~\tev$)  
      \\
      \hline
      \hline
    \end{tabular}
    \end{equation}
%
\item[LEP Searches] 
    The Large Electron-Positron Collider (LEP) performed searches for light BSM Higgs bosons both using the $e^+e^-\to HZ$ and $e^+e^- \to AH$ channel~\cite{Schael:2006cr}. The $HZ$ production rate scales proportionally to $\cba$ and the null results of this search can constrain $\cba$ down to about 0.1 for $m_H$ between 12.5 to 114.4 GeV. This constraint does not apply to any of the benchmarks considered in this paper.   In contrast, the double Higgs production channel $AH$ is unsuppressed under the alignment limit and roughly constrains $m_A + m_H \lesssim 200~\gev$.
\item[Measurements of SM processes]
    In the absence of dedicated resonance searches, additional BSM Higgs states can also be probed through inclusive cross section measurements of rare SM processes. In this work, we consider two such processes: i) The exotic Higgs decay $A/H \to HZ/AZ\to ttZ$, which dominates for daughter Higgs masses above the di-top threshold, can be probed using the $ttZ$ cross section measurement~\cite{Aaboud:2019njj}. ii) The associated $ttH/ttA$ production channel with subsequent decay $H/A\to tt$ is sensitive to the $4t$ production rate \cite{Sirunyan:2019wxt, Aaboud:2018jsj,Aaboud:2018xpj}. This search constrains low $\tb<1$ for $m_{A/H}>2m_t$, where the Higgs width becomes very large, $\Gamma/m>0.2$ and hence no resonance search can be performed anymore. 

\end{description}

While the above constraints focus on the neutral Higgs bosons, additional theoretical and experimental constraints arise for  the charged Higgs bosons $H^\pm$.  
\begin{description}
\item [Precision Constraints]    
    Electroweak precision measurements~\cite{ALEPH:2005ab,Haller:2018nnx} require the charged Higgs mass to be close to the mass of one of the neutral Higgses, $m_{H^\pm}\sim m_H$ or $m_A$ or $m_h$~\cite{Kling:2016opi}.
\item [Direct Searches]
    Searches for charged Higgses $H^\pm$ have been performed through the decay channels $H^\pm \to cs$~\cite{Aad:2013hla, CMS:2019sxh}, $\tau \nu$~\cite{Aaboud:2018gjj, Sirunyan:2019hkq} and $t b$~\cite{Aaboud:2018cwk, Sirunyan:2019arl}. At large masses of $m_{H^\pm} > m_t$, constraints on these charged Higgs searches are typically weaker than their neutral counterparts, due to the suppressed $tbH^\pm$ associated production cross section as well as large backgrounds. A notable exception are searches of light charged Higgs bosons via top decays $t \to H^\pm b$, which exclude the mass range $m_{H^\pm} \lesssim m_t$ for the Type-II 2HDM and part of the parameter space for the Type-I 2HDM.
\item [Flavour Constraints] 
    Precision flavour observables, such as of the branching fraction of $B$ mesons, provide indirect constraints on the charged scalars. Most notably, the measured value for BR($b \to s \gamma$) imposes a lower limit on the charged Higgs mass to be larger than  $\sim 600~\gev$ in the Type-II 2HDM~\cite{Haller:2018nnx,Amhis:2016xyh}. However, we note that the interpretation of these flavour measurements are typically model dependent and contributions from additional BSM sectors could significantly modify and relax these constraints. 
\end{description}
Given the above considerations, we do not consider any additional constraints related to the charged Higgs bosons and solely focus on the neutral Higgs sector in this paper.

\section{Degenerate Higgs Masses}
\label{sec:degenerate}

\begin{figure*}[t]
\centering
\includegraphics[width=0.49\textwidth]{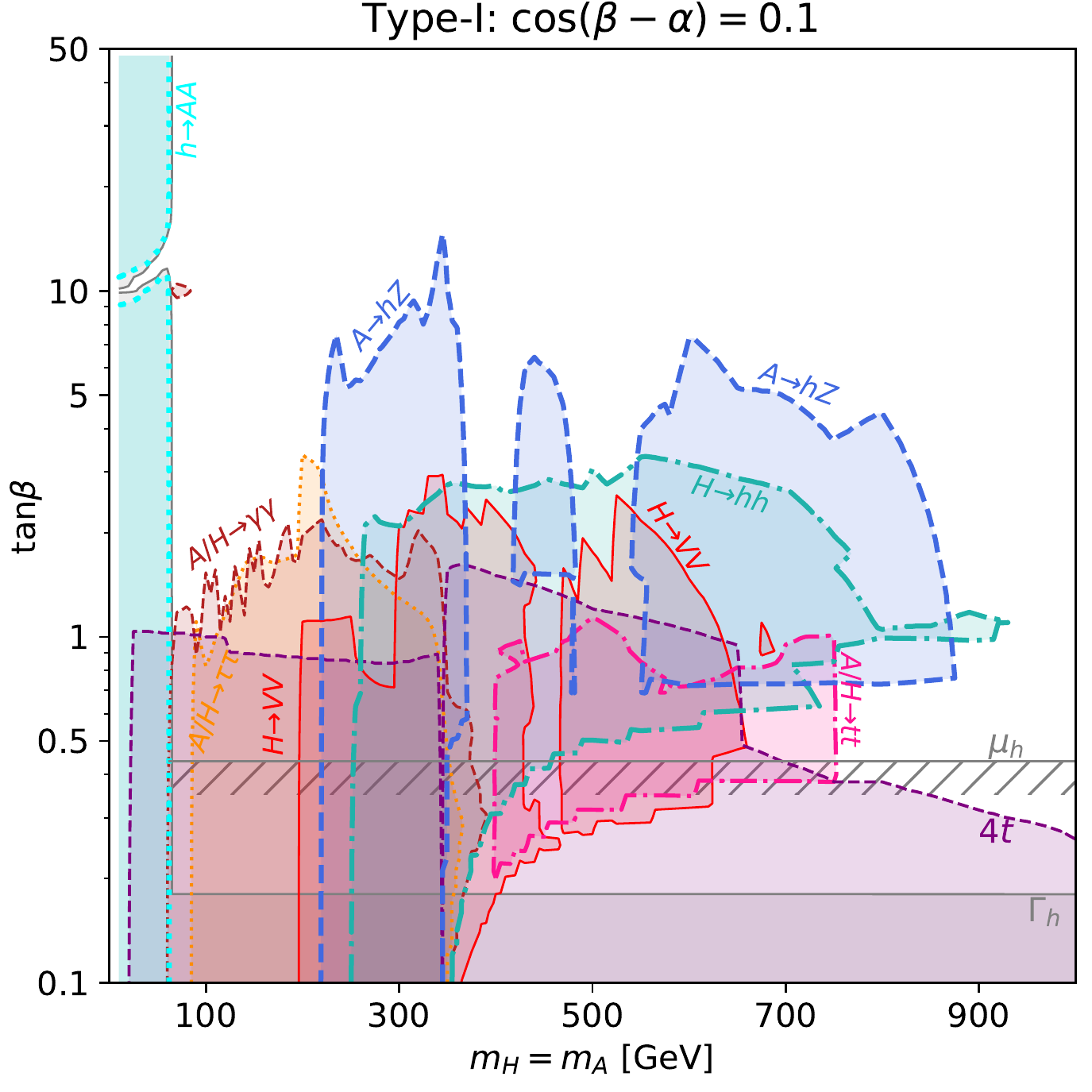}
\includegraphics[width=0.49\textwidth]{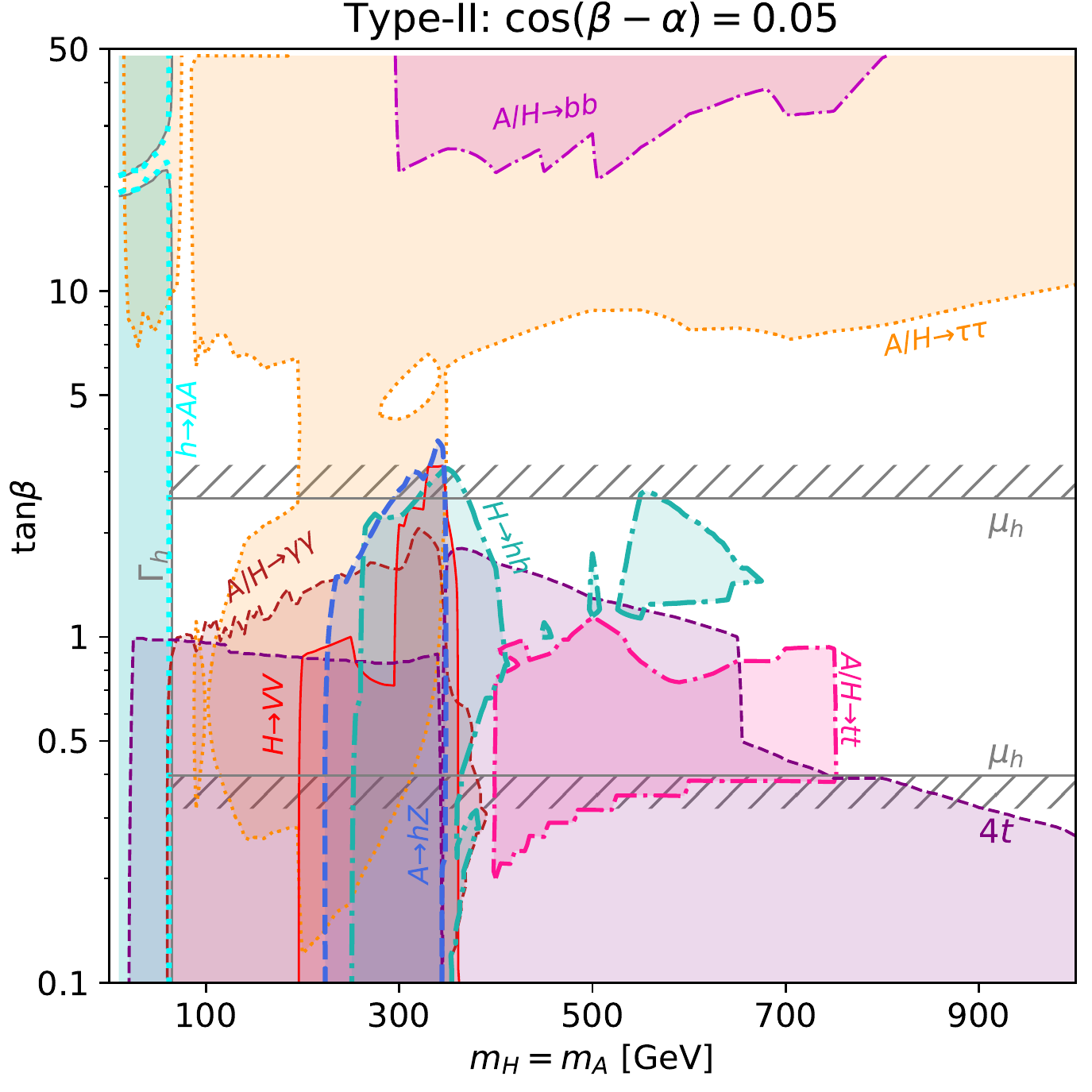}
\caption{
Constraints for degenerate heavy Higgs mass spectrum $m_A=m_H=m_{H^+}$. We show the 95\% C.L. exclusion region in the $m_{A/H}$~vs.~$\tb$ plane on the Type-I 2HDM with $\cba=0.1$ (left) and Type-II 2HDM with $\cba=0.05$ (right) originating from i) the measurement of the Higgs width $\Gamma_h$ (grey), ii) the conventional search results on $H/A \to \tau\tau$ (dotted orange), $H/A \to bb$ (dot-dashed pink), $H \to VV$ (red), $H/A \to \gamma\gamma$ (dashed brown), $H/A \to tt$ (dot-dashed magenta) and $4t$ production (dashed purple), and iii) exotic decay channels $A\to hZ$ (dashed dark blue), $H\to hh$ (dot-dashed green) and $h \to AA$ (dotted cyan). Region enclosed by the grey hatched line are excluded at 95\% CL. by the current Higgs coupling measurements. 
}
\label{fig:conven} 
\end{figure*}

To interpret the experimental results, we take the limits on cross section times branching fraction $\sigma \times \text{BR}$   of the  various channels mentioned above, as well as  the measurements of Higgs couplings and $\Gamma_h$ to directly constrain the 2HDM parameter space.  We use the {\tt SusHi} package \cite{Liebler:2016ceh}  to calculate the production cross-section at NNLO level, and  the {\tt 2HDMC}~\cite{Eriksson:2009ws} code for Higgs decay branching fractions at tree level.   
In \autoref{fig:conven}, we show the current collider limits in the 2HDM $m_{H/A} - \tb$ plane, taking into account the  constraints mentioned above.  We assume degenerate heavy Higgs masses $m_A=m_H=m_{H^+}$ such that exotic decays  involving two BSM Higgses are not kinematically open.   The mixing angle $\cba$ is fixed to be 0.1 for the Type-I 2HDM (left panel) and 0.05 for the Type-II 2HDM (right panel), based on the consideration of the SM-like Higgs coupling measurements.  Non-zero values of $\cba$ are relevant for $A\to Zh$, $H \to hh$ and $H\to VV$.  The soft $\mathcal{Z}_2$ breaking parameter $m_{12}^2$ is chosen to be $ m_{12}^2=m_H^2 s_\beta c_\beta$ to satisfy theoretical consideration of unitarity and vacuum stability. \medskip

For the Type-I 2HDM (left panel), current direct searches are mostly sensitive at $\tb<10$. This is because the main production modes, gluon fusion and $b$-associated production, are both $\cot \beta$-enhanced from bottom and top Yukawa couplings. The limits from the  conventional modes of $A/H \to\tau\tau$ (orange)  and $A/H \to \gamma\gamma$ (brown) have weak dependence on the value of $\cba$ and exclude the low mass region below the top threshold of $m_{A/H}\lesssim 2 m_t$ for $\tb < 3$. Once  the $tt$ mode (magenta) is open, it quickly dominates the decay branching fractions.   The region of $400\ \gev< m_{A/H}<  750\ \gev$ with $0.2 <\tb< 1$ is currently excluded by this channel.  For even smaller $\tb$, no limits are quoted for the $tt$ channel because the corresponding Higgs width is so wide that the resonant search results are not applicable~\cite{Sirunyan:2019wph,Aaboud:2019roo}.   

The limits for $H\to VV$ (red), $A\to Zh$ (blue), and $H\to hh$ (green)  strongly depend on the value of $\cba$  and vanish under the alignment limit of $\cba=0$.  For $\cba=0.1$, $m_{A/H}$ between 200 and 850 GeV for $0.5 <\tb<10$ are excluded, with a gap  at intermediate masses around 450 GeV for $A \to Zh$ and $H \to VV$ channel.   

The 95\% C.L. range of the SM-like Higgs decay width (grey)  excludes the low mass region of $A/H$ given the opening of $h \to AA,\ HH$ for $m_{A/H} < m_h/2$, as well as low $\tb$ region of $\tb \lesssim 0.2$ for $\cba=0.1$ due to the enhancement of fermion Yukawa couplings.  A thin slice of surviving region from $\Gamma_h$ constraints around $\tb\sim 10$  remains due to the vanishing of $\Gamma(h\to AA)$ in that region.   Additionally, a global fit to the LHC SM-like Higgs coupling measurements excludes $\tb \lesssim 0.4$ for $\cba=0.1$.   Finally, the measurement of the four top production (purple) rate is sensitive to $ttH/ttA$ associated production with $A/H\to tt$,  constraining a  wide region at low $\tb$.  

For the Type-II 2HDM (right panel), the results are quite different at large $\tb$  due to the $\tb$-enhanced bottom and lepton Yukawa couplings. At large $\tb$, the  $A/H \to \tau\tau$ provides the strongest constraints, excluding $\tb\gtrsim 10$ for a large range of BSM Higgs masses. Below the top threshold, $100~\gev < m_H<2m_t$, this channel can nearly probe the entire range of $\tb$, setting the strongest constraints~\cite{Aad:2020zxo}. The exclusion region for $\gamma\gamma$ is similar to those of the Type-I 2HDM, while  the exclusion region for $VV$, $H\to hZ$ and $H \to hh$ are reduced comparing to the left panel, given the smaller value of $\cba$ used here. The global fit to the LHC SM-like Higgs coupling measurements excludes $\tb \lesssim 0.4$  and $\tb \gtrsim 2.5$ for $\cba=0.05$. 

\section{Non-degenerate Case:  $H/A \to AZ/HZ$ }
\label{sec:exotic}

The previous section focused on the degenerate case of $m_H = m_A = m_{H^\pm}$.  Once there is a sizeable mass splitting between the BSM Higgs masses, additional exotic channels such as $H/A \to AZ/HZ/H^\pm W^\mp $ will open up and quickly dominate the decay branching fractions.   As a result, the reach of conventional searches shown in the last section will be reduced.  At the same time, these exotic channels provide new opportunities for BSM Higgs searches at the LHC~\cite{Coleppa:2013xfa,Coleppa:2014cca,Coleppa:2014hxa,Hajer:2015eoa,Kling:2015uba,Kling:2018xud}.

Under the alignment limit, the decay branching fractions of $H/A \to AZ/HZ/H^\pm W^\mp $ are unsuppressed. The most promising final states are $A/H \to HZ/AZ \to bb\ell\ell$ and $\tau\tau\ell\ell$, which allow for a clean identification through the dileptons from $Z$ decay~\cite{Coleppa:2013xfa,Coleppa:2014hxa}. These modes have been analysed by both CMS~\cite{Khachatryan:2016are, Sirunyan:2019wrn} and ATLAS \cite{Aaboud:2018eoy}, as listed in 
~\autoref{sec:conven}.

In what follows, we will first present the constraints on the parameter space of $\{\tb, \cba, m_A, m_H\}$ from $H/A \to AZ/HZ$ channel alone, focusing on the parameter dependence.  We will then present $H/A \to AZ/HZ$ together with all other direct and indirect search channels and discuss the complementarity of various channels.
  
\subsection{$m_A$ vs. $m_H$}

\begin{figure*}[t]
\centering
\includegraphics[width=0.49\textwidth]{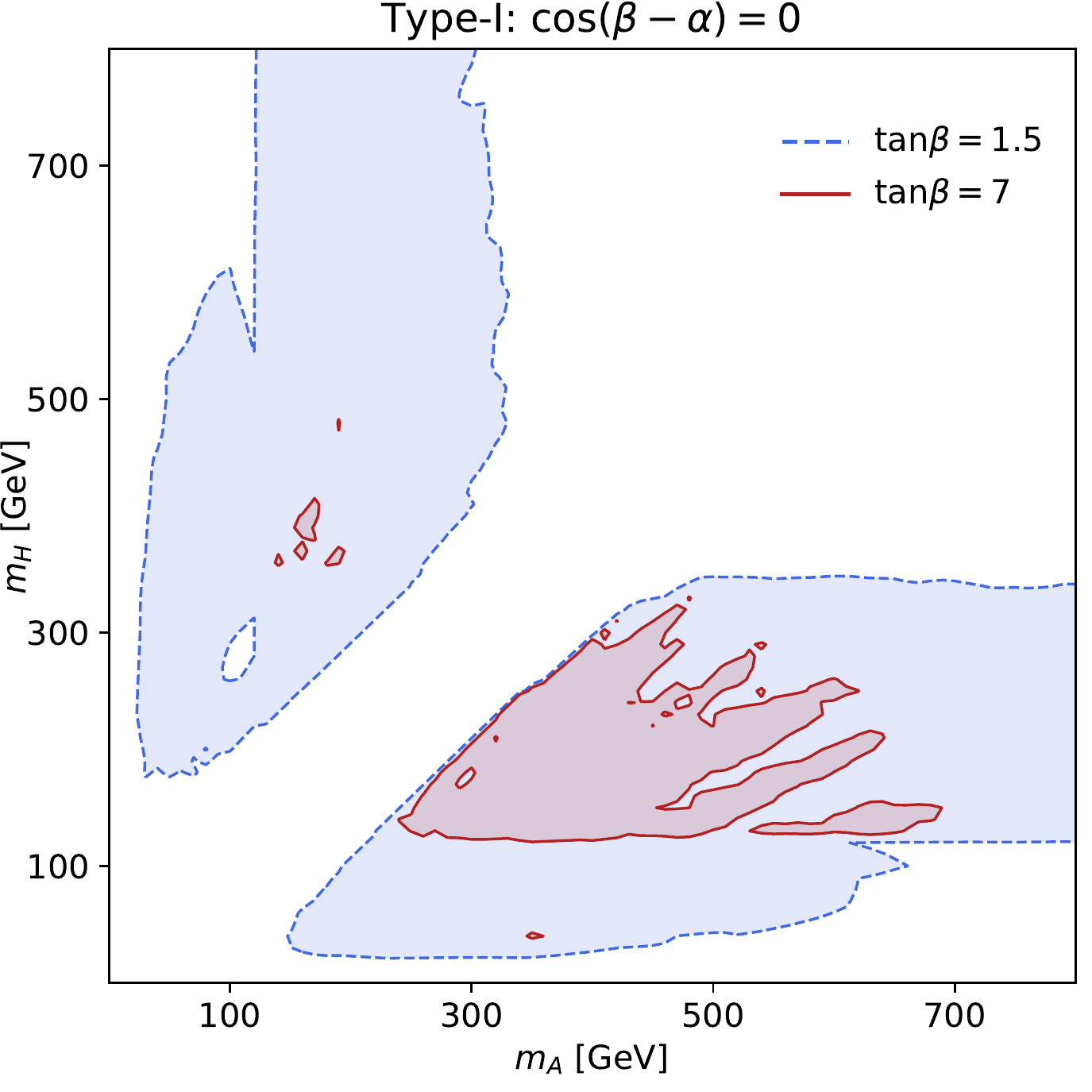}
\includegraphics[width=0.49\textwidth]{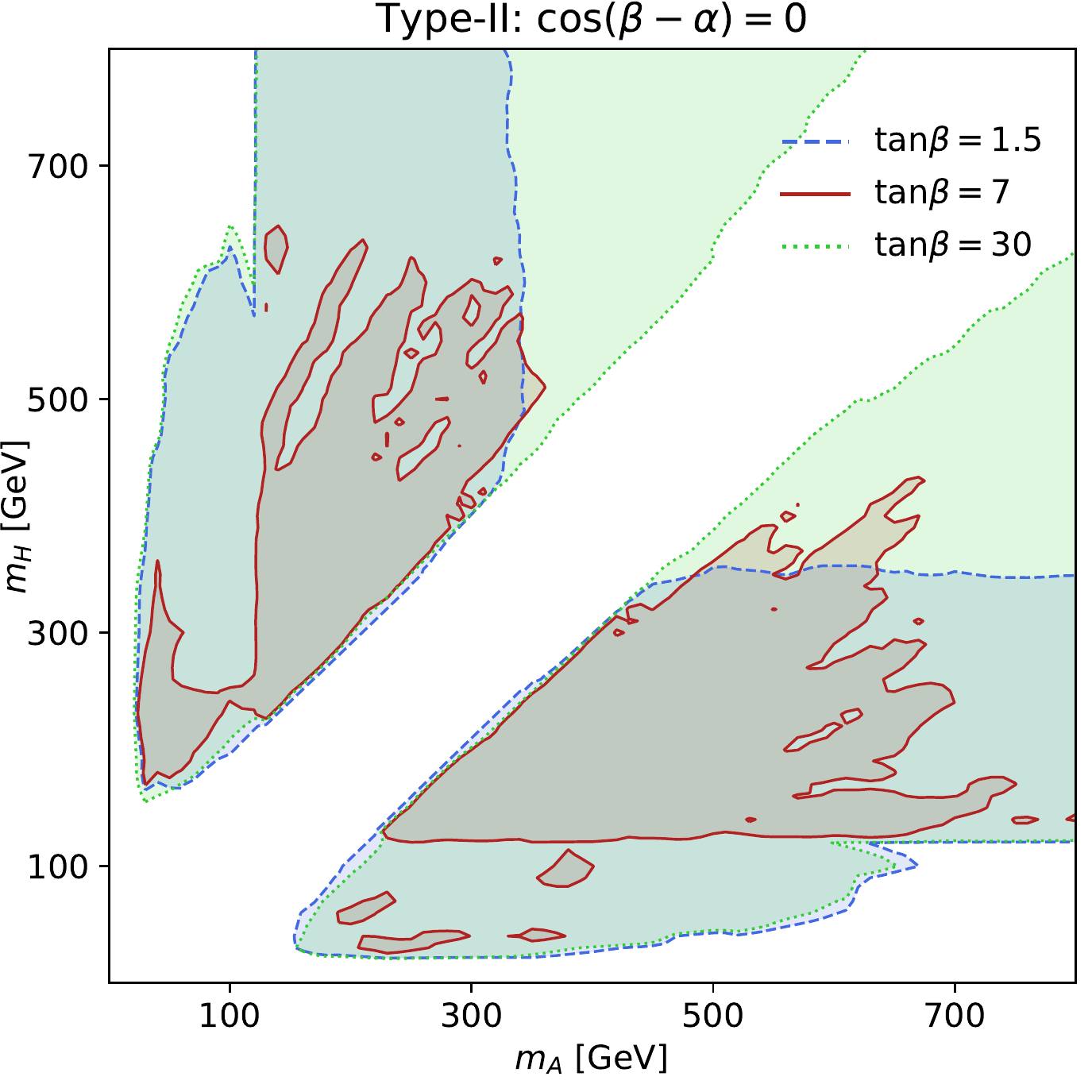}
\includegraphics[width=0.49\textwidth]{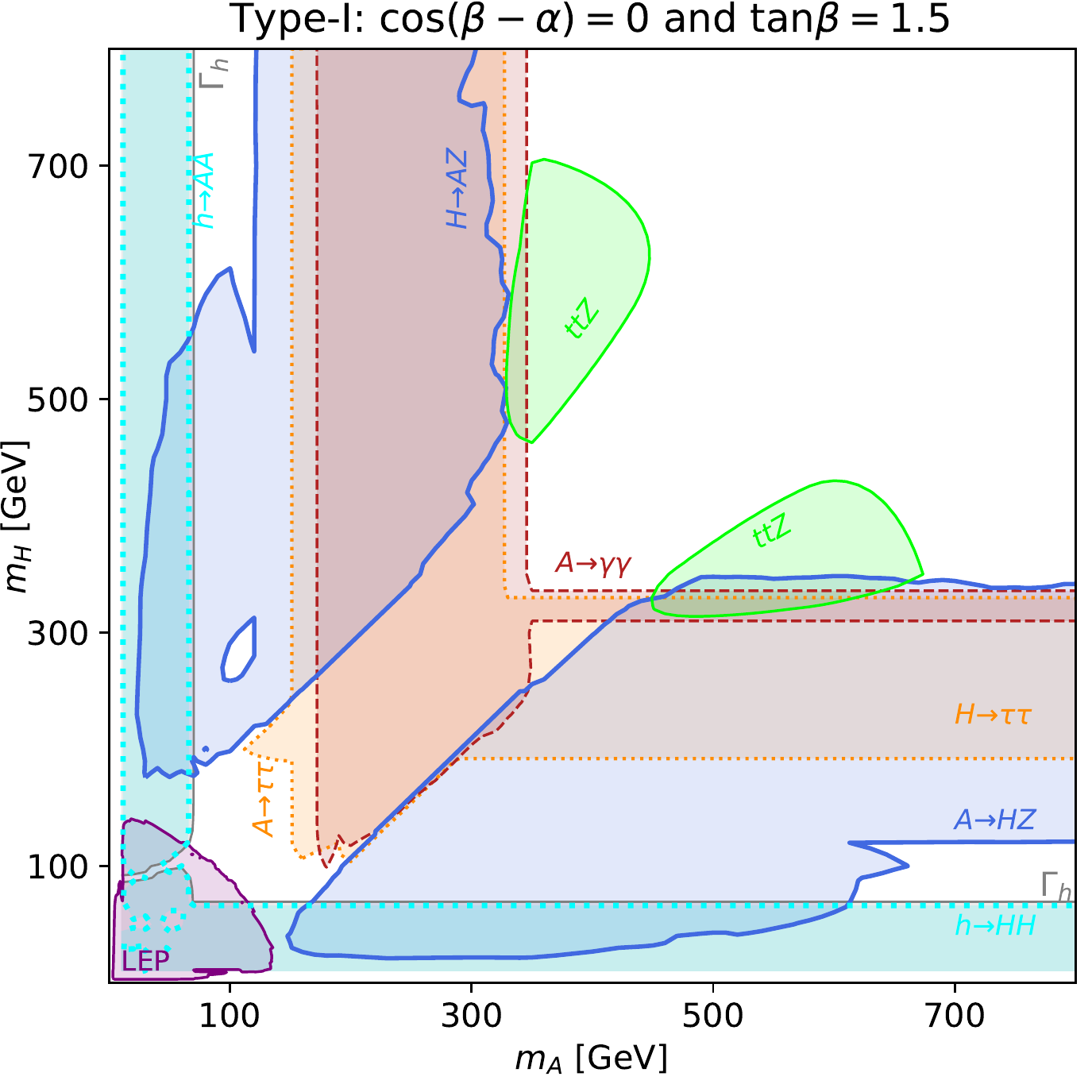}
\includegraphics[width=0.49\textwidth]{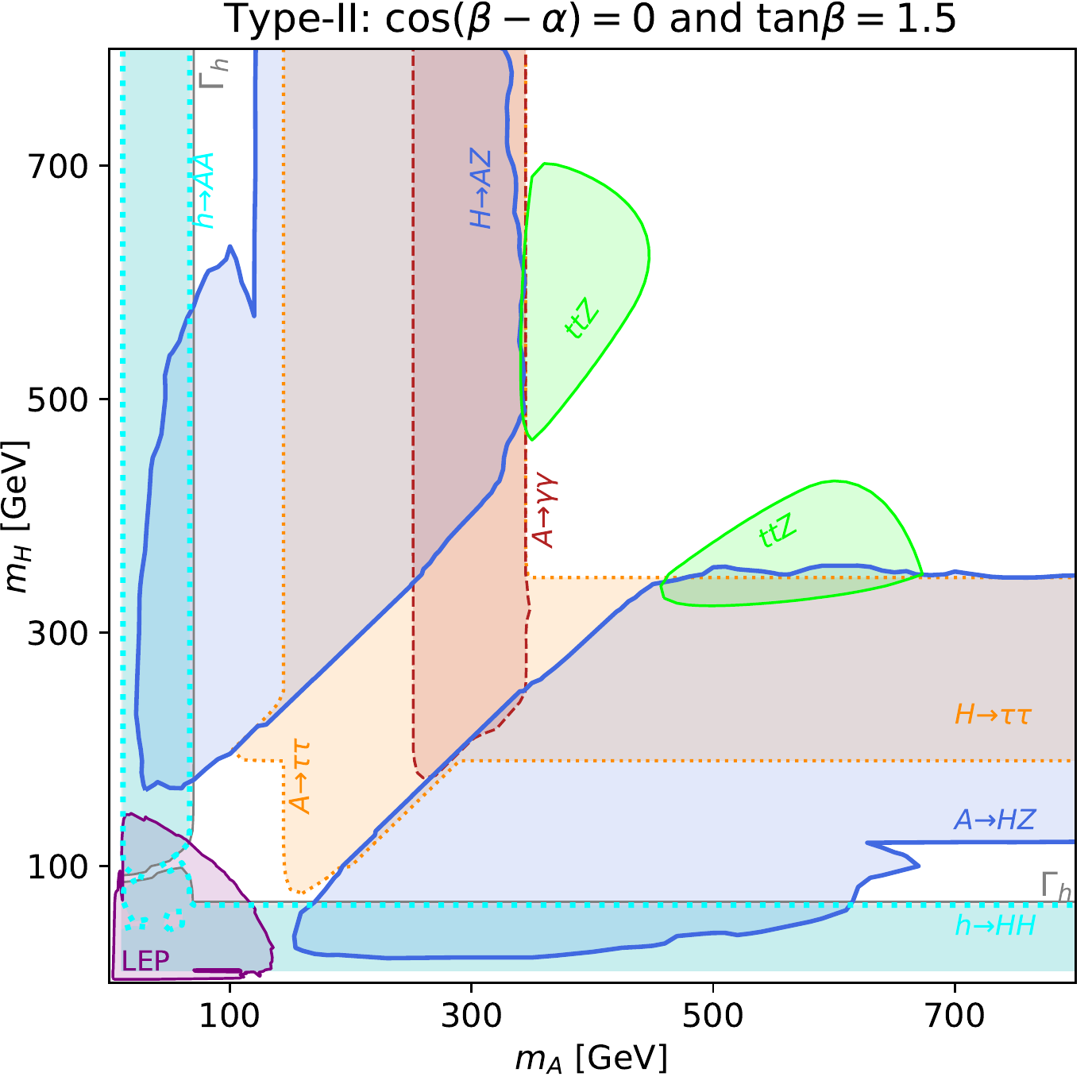}
\caption{
Constraints on the Type-I (left panel) and the Type-II (right panel) 2HDM in $m_A$ vs. $m_H$ plane. 
\textbf{Top:} Parameter space excluded at 95\% C.L. by the $A/H \to HZ/AZ$ search in the alignment limit, $\cba=0$, for $\tb=1.5$ (blue), $7$ (red) and $30$ (green). 
\textbf{Bottom:} Constraints at 95\% C.L. for $\cba=0$ and $\tb=1.5$ from LHC searches for $A/H \to HZ/AZ$ (blue), $A/H \to \tau\tau$ (dotted orange), $A/H \to \gamma\gamma $ (dashed brown), $h\to AA/HH $ (dotted cyan) and $ttZ$ production (green) as well as LEP searches (purple) and the Higgs width measurement $\Gamma_h \in (0.08,9.16)~\mev$ (grey). 
}
\label{fig:mAmH}
\end{figure*}

Let us study the explicit heavy Higgs mass dependence by looking at the $m_A$ vs. $m_H$ plane of the 2HDM parameter space. In the top panel of \autoref{fig:mAmH} we show the constraints from the $A/H \to HZ/AZ$ channel for the Type-I (left) and Type-II (right) in the alignment limit, $\cba=0$. Away from the alignment limit, these constraints are weakened given the suppressed coupling $g_{HAZ}\propto\sba$. In the gap region along $m_A\sim m_H$, the exotic decay modes are kinematically inaccessible. 

For the Type-I 2HDM with low $\tb=1.5$ (blue), the 13 TeV searches exclude parent particle masses up to 800 GeV for a daughter particle mass between 80 and 350 GeV. For higher values of $\tb$, these constraints are weakened due to the suppression of Yukawa couplings $g_{Aff/Hff}\sim \tb^{-1}$ resulting in a reduced production cross section. The reach for intermediate $\tb=7$ region (red) is reduced greatly, while it vanishes for even larger values of $\tb$. The asymmetry between $A$ and $H$ is due to the different parent particle production cross sections as well as daughter particle decay branching ratios. 

While at low $\tb=1.5$ the reach for the Type-I and Type-II 2HDM are very similar, at large $\tb$ the Type-II 2HDM has an enhanced reach due to the $\tb$ enhancement of bottom (and $\tau$) Yukawa couplings. At $\tb=30$ (green) the $A/H \to HZ/AZ$ search channel constrains the kinematically allowed region up to Higgs masses of $m_{A/H} \sim 800~\gev$, with the exception of very small daughter particle masses. The constraints are weakest for intermediate values of $\tb \sim 7$ case (red), with the parent particle mass excluded only up to about 700 GeV.

\medskip  
  
In the lower panels of \autoref{fig:mAmH} we present the global constraints on the 2HDM parameter space for $\cba=0$ and $\tb=1.5$. In particular, we show the regions excluded by Higgs searches via the $A/H \to HZ/AZ$ (blue), $A/H \to \tau\tau$ (orange), $A/H \to \gamma\gamma $ (brown) and $h\to AA/HH$ (cyan) channels as well as $ttZ$ rate measurements (green). Furthermore we include the LEP search results (purple) and constraints arising from the measurements of the Higgs width, $\Gamma_h \in (0.08,9.16)~\mev$ (grey). Additional constraints from the $A\to Zh$ and $H\to VV,~hh$ channels vanish in the alignment limit, while search results from $A/H \to bb$, $\mu\mu$ and $tt$ are too weak to set any constraint.   

For both Type-I and Type-II 2HDMs, the combination of all channels cover the majority of the region in which one of the Higgs masses is below the di-top threshold, $m_{A},m_H < 2m_t$. In addition to $A/H \to HZ/AZ$, these constraints come from direct searches for the lighter BSM Higgs state which decays into conventional final states $A/H \to \gamma\gamma$ and $\tau\tau$. In particular, the gap region for $A/H \to HZ/AZ$ is mostly covered by these searches. However, these channels become inefficient for Higgs mass above $2m_t$, where $A/H \to tt$ opens up. This not only decreases the branching fraction into the clean $A/H \to \gamma\gamma$ and $\tau\tau$ final states but it can also increase the heavy Higgs widths significantly which imposes a general problem for resonant searches. 

Interestingly, this region can be probed by the measurements of $ttZ$ rate, which effectively constraints the process $A/H \to HZ/AZ \to ttZ$. Parent particle masses around $450-700~\gev$ and daughter particle mass around $350-450~\gev$ are excluded by this rate measurement. Although a resonant search would be challenging due to large Higgs widths, this result motivates a dedicated Higgs search utilizing this channel. 
 
At very low masses, $m_A,m_H<m_h/2$, the BSM Higgs states can be produced in the decay of the SM Higgs $h\to AA,HH$. These channels have been constrained both directly in a variety of final states outlined in \autoref{sec:conven} and via their impact on the Higgs width itself. Both direct searches and indirect Higgs width measurements result in similar constraints in the 2HDM parameter space, excluding light masses $m_A,m_H<m_h/2$. Finally, searches at LEP provide additional constraints at low masses $m_A+m_H\lesssim 200\ \gev$.
 
\subsection{$\tan\beta$ vs. $\cos(\beta-\alpha)$}

\begin{figure*}[tbh]
\centering
\includegraphics[width=0.49\textwidth]{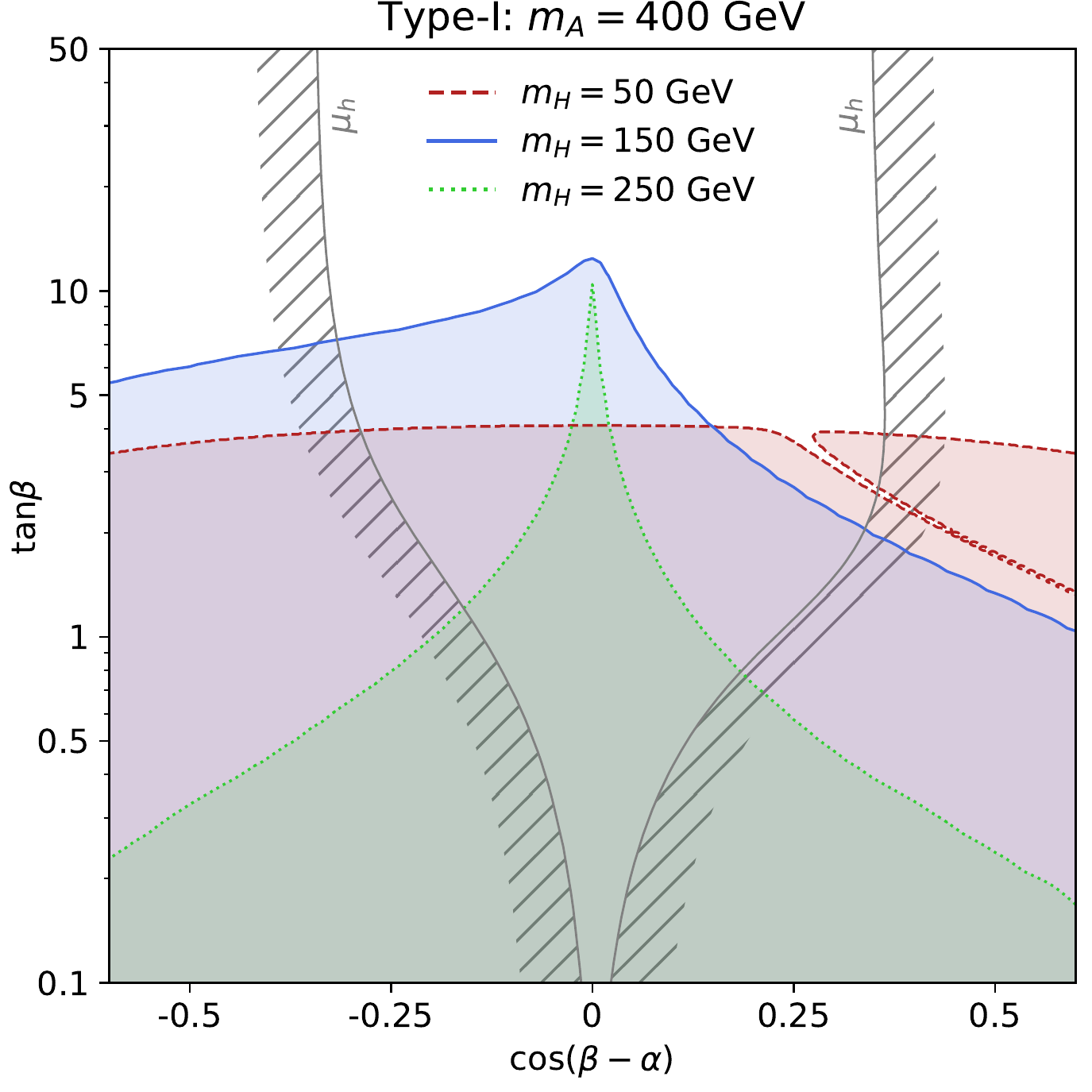}
\includegraphics[width=0.49\textwidth]{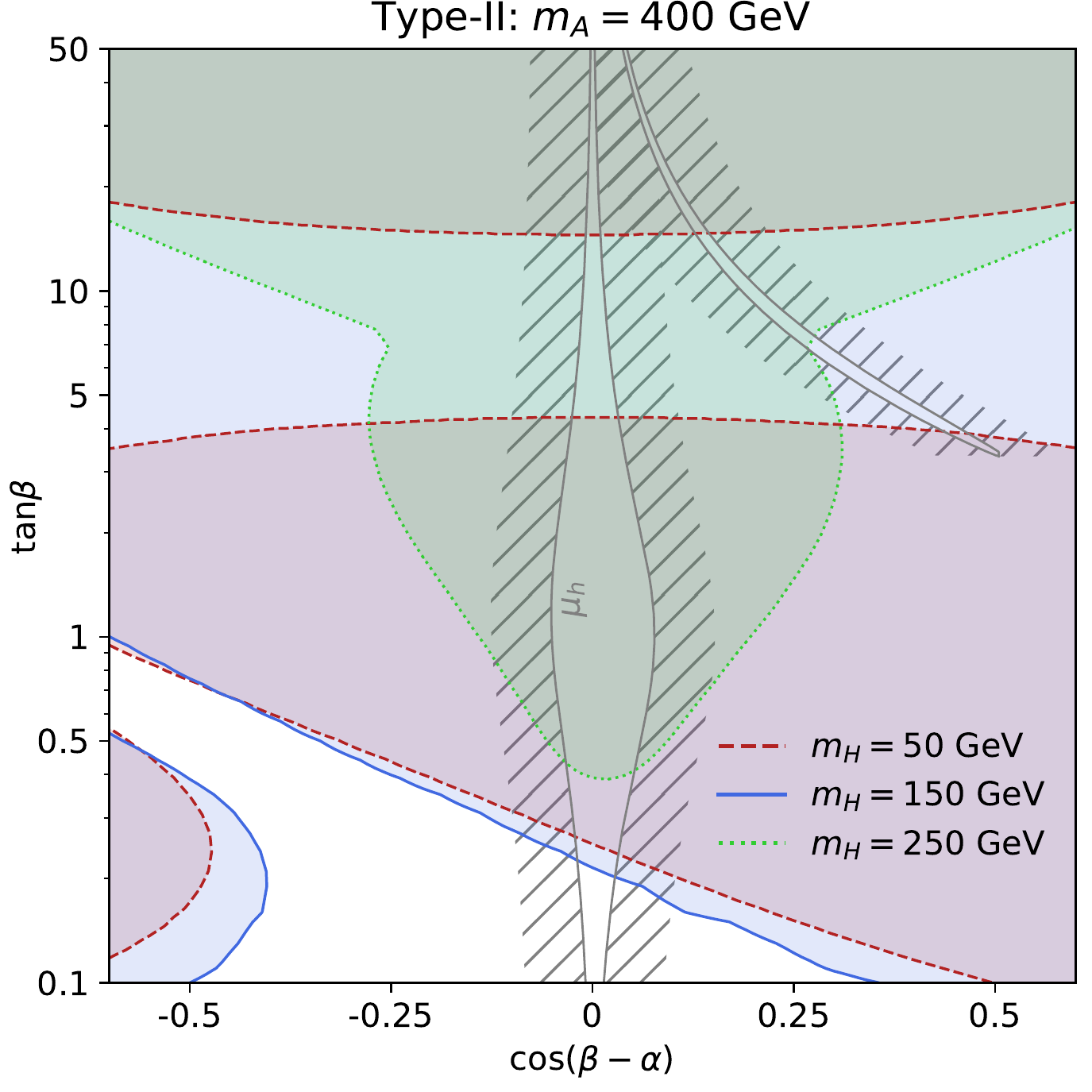} 
\includegraphics[width=0.49\textwidth]{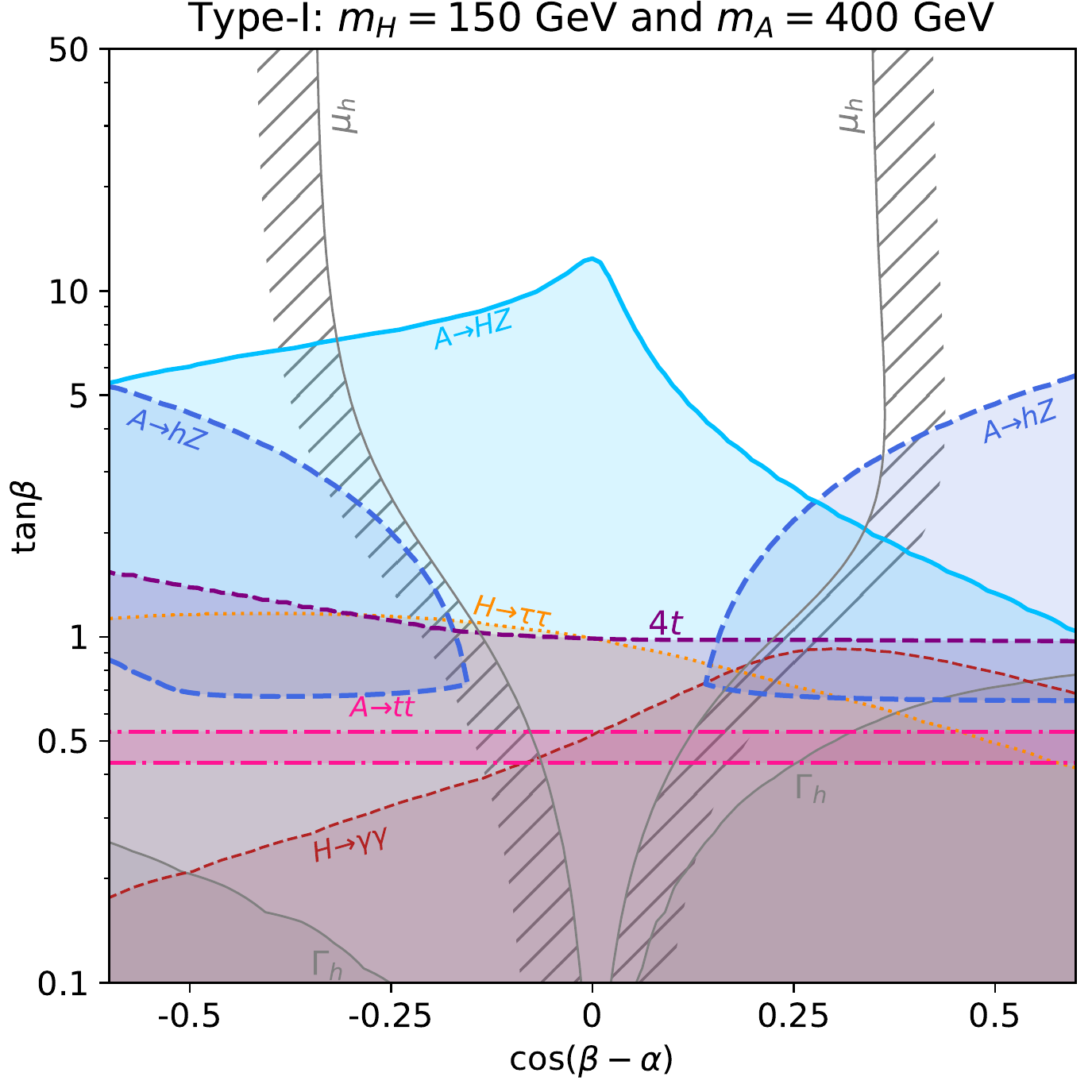}
\includegraphics[width=0.49\textwidth]{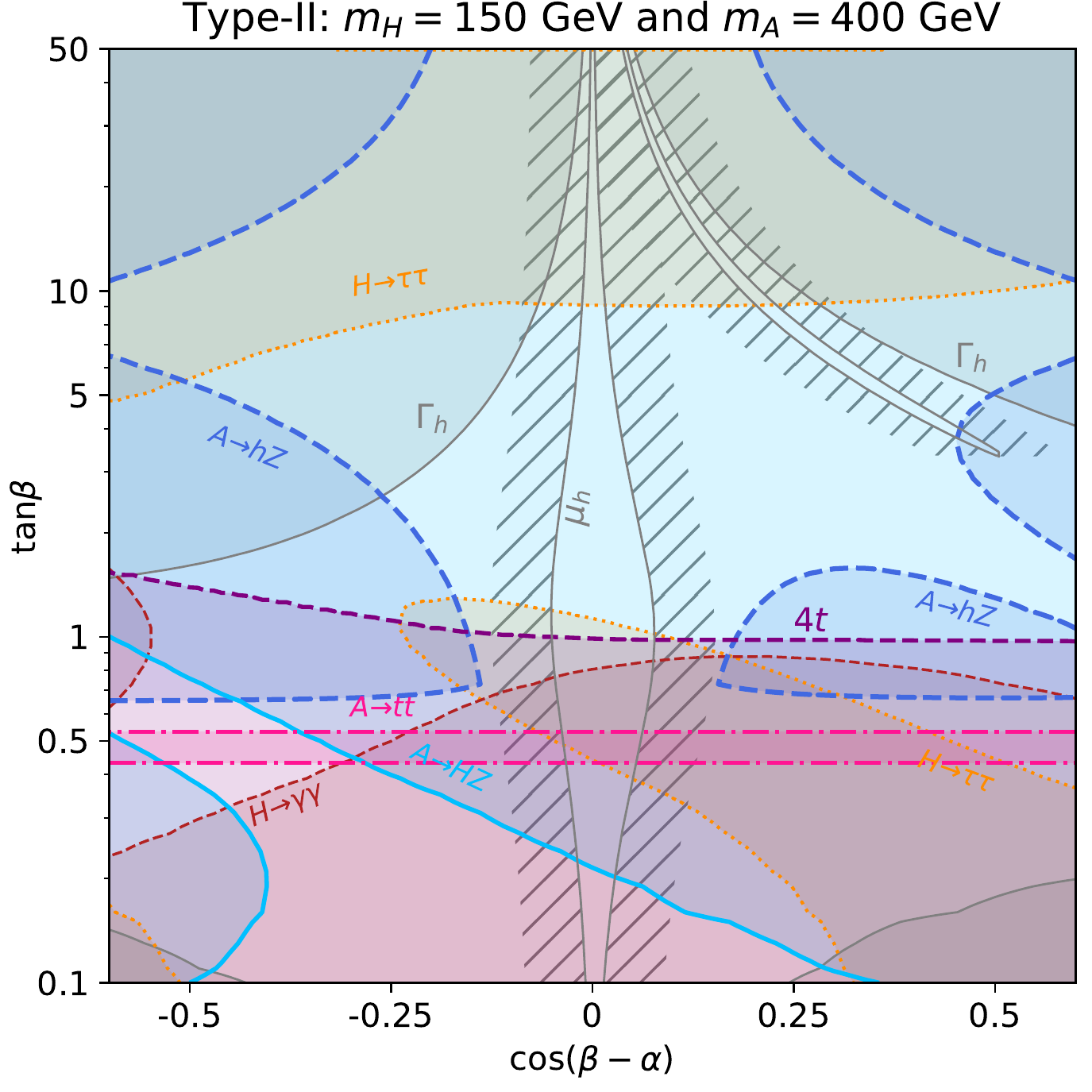}
\caption{
Constraints on the Type-I (left panel) and the Type-II (right panel) 2HDM in $\tb - \cba$ plane. 
\textbf{Top:} Parameter space excluded at 95\% C.L. by the $A/H \to HZ/AZ$ search for $m_A=400~\gev$ and $m_H=50~\gev$ (red), $150~\gev$ (blue) and $250~\gev$ (green), and by the global fit of SM-like Higgs couplings strength $\mu_h$ (grey).
\textbf{Bottom:} Constraints at 95\% C.L. for $m_A=400~\gev$ and $m_H=150~\gev$ from LHC searches for $A \to HZ$ (blue), $A\to hZ$ (dashed dark blue), $H \to \tau\tau$ (dotted orange), $H \to \gamma\gamma$ (dashed brown), $A \to tt$ (dot-dashed magenta) and $4t$ production (dashed purple) as well as the global fit of SM-like Higgs couplings strength $\mu_h$ (grey hatched region).  
}
\label{fig:cbatb}
\end{figure*}

Let us now compare the reach of the direct BSM Higgs searches with the indirect constraints from the precision SM-like Higgs coupling measurements, which are best studied in the $\cba$ vs. $\tb$ plane. 

In \autoref{fig:cbatb}, we show the constraints from the latest LHC Higgs coupling precision measurements as the grey shaded region for both Type-I (left) and Type-II (right) 2HDMs. The central region around the alignment limit of $\cba=0$ is always unconstrained by the Higgs coupling measurements. For the Type-I 2HDM, the fermion Yukawa couplings can be written as $g_{hff} = \sba + \cba / \tb$. At low $\tb$, even small deviations from the alignment limit results in an enhanced Higgs production rate and decay width into bottom/tau pairs which can be excluded by Higgs couplings measurements. For large values of $\tb$, the Yukawa couplings are almost $\tb$-independent, $g_{hff} \sim \sba$, resulting in weaker constraints of $|\cba|<0.35$, mainly from the measurement of the vector boson couplings $g_{hZZ}$. 

For the Type-II 2HDM, both the small and large $\tb$ region are tightly constrained by fermion Yukawa couplings. Away from the alignment limit, the SM-like Higgs production rate in gluon-gluon fusion quickly increases as low $\tb$, while the Higgs decay widths into bottom and tau pairs are enhanced at high $\tb$.  The bounds are weakest for intermediate values of $\tb\sim 1$, constraining $|\cba|<0.08$. Once loop contributions from heavy Higgses are included, the shape of the allowed region will be tilted, with tighter constraints at the large $\tb$ region for the Type-I 2HDM, as shown in Refs.~\cite{Chen:2018shg, Gu:2017ckc, Chen:2019pkq}. For the Type-II 2HDM, there is an additional allowed wrong-sign Yukawa ``arm" region~\cite{Ferreira:2014naa, Gu:2017ckc, Su:2019ibd, Han:2020zqg} for $0.1 < \cba <0.5 $ and $\tb > 4$. 
\medskip

The upper panels of \autoref{fig:cbatb} also show the bounds obtained from the exotic Higgs decay searches for $A\to HZ \to bb\ell\ell/\tau\tau \ell\ell$ for fixed $m_A=400~\gev$ and for $m_H=50~\gev$ (red), $m_H=150~\gev$ (blue) and $m_H=250~\gev$ (green). For the Type-I 2HDM (left), small $\tb \lesssim 5$ region is excluded. For $m_H=50~\gev$, the bounds are almost independent of $\cba$, with the exception of a small band around $g_{Hff} = \cba - \sba/\tb = 0$, where the decay $H \to bb/\tau\tau$ is suppressed and the sensitivity vanishes. For larger daughter masses, the sensitivity decreases away from the alignment limit, since the the $H \to bb/\tau\tau$ branching fraction is reduced both due to the suppression around $g_{Hff}=0$ and the increasing decay branching fractions for the competing $H \to VV$ and $hZ$ channels at large $|\cba|$.  The later is particularly important for heavy Higgs $m_H > 2 m_W$, where the  sensitivity sharply peaks around the alignment limit, as can be seen from the green curve for $m_H=200~\gev$. 

The Type-II 2HDM (right panel), on the other hand, also receives constraints at large $\tb$ due to the enhanced bottom-associated Higgs production mode. For $m_H=50$ and $250~\gev$, one can identify two distinct regions corresponding to $bbA$ associated production at high $\tb \gtrsim 10$ and the gluon fusion at low $\tb \lesssim 10$, while for $m_H=150$ both these regions overlap, covering the entire parameter space. As for Type-I, the sensitivity decreases for large $|\cba|$ due to the competing $H \to VV$ and $hZ$ channels, which is especially visible for $m_H=250~\gev$. A band with no sensitivity is now visible around $g_{Hbb}=g_{H\tau\tau} = \cba + \sba \tb = 0 $ as a result of the cancellation between the first and second term.  \medskip

The lower panels of \autoref{fig:cbatb} show the complementary between the Higgs couplings precision measurements (grey hatched region) and the direct exclusion limits from various BSM Higgs search channels for $m_A=400~\gev$ and $m_H=150~\gev$. The leading constraints come from the heavy CP-even Higgs decay $H \to \tau\tau$ and $\gamma\gamma$, the CP-odd Higgs decay $A \to tt$ and $hZ$, as well as the $4t$ cross section measurement.  We do not show additional constraints from the  $H/A \to VV, hh, \mu\mu$ and $bb$ channels, which are generally weaker. 

For Type-I, the $H \to \tau\tau$ and $\gamma\gamma$ channel constrain $\tb<1$ with a $\cba$ dependence entering both in the production and decay. The $A \to tt$ constraints are independent of $\cba$ and limited by a suppressed production rate towards larger $\tb$ and a large decay width $\Gamma_A > 0.2 m_A$ at low $\tb$. The $A \to hZ$ channel constrains $|\cba|>0.2$ for $0.8\lesssim\tb\lesssim 5$, where the low $\tb$ limit is again due to a large width $\Gamma_A$, limiting the applicability of the resonance search. Finally the $4t$ cross section measurement can constrain the $\tb \lesssim 1$ region, with the limits constraining the (offshell) $ttH$ production at negative $\cba$ and $ttA$ production positive $\cba$, resulting in $\cba$-dependent and $\cba$-independent bounds, respectively. 

For Type-II, the additional $bbA/bbH$ production modes result in additional constrains from $H\to \tau\tau$ and $A \to hZ$ at high $\tb>10$. Additionally, the $H\to \tau\tau$ reach is reduced at low $\tb<1$ for negative $\cba$ where the $g_{H\tau\tau}=\cba + \sba \tb$ coupling vanishes. Also note the exclusion region of $A \to hZ$ for $\cba>0$ and intermediate $\tb \sim 2$ is split in two parts, due to the vanishing branching branching fraction for $h \to bb$ when $g_{hbb} = \sba - \cba \tb = 0 $. 

Combining all constraints, we can see that the Higgs precision measurements exclude large deviations away from the alignment limit, while the direct   BSM Higgs searches provide additional constraints also in the alignment limit. In particular, the exotic Higgs decay $A \to HZ$ provides important constraints for intermediate values of $\tb$, showing its complementarity to the conventional searches. 

\subsection{$\tan\beta$ vs. $m_A$}

\begin{figure*}[tbh]
\centering
\includegraphics[width=0.49\textwidth]{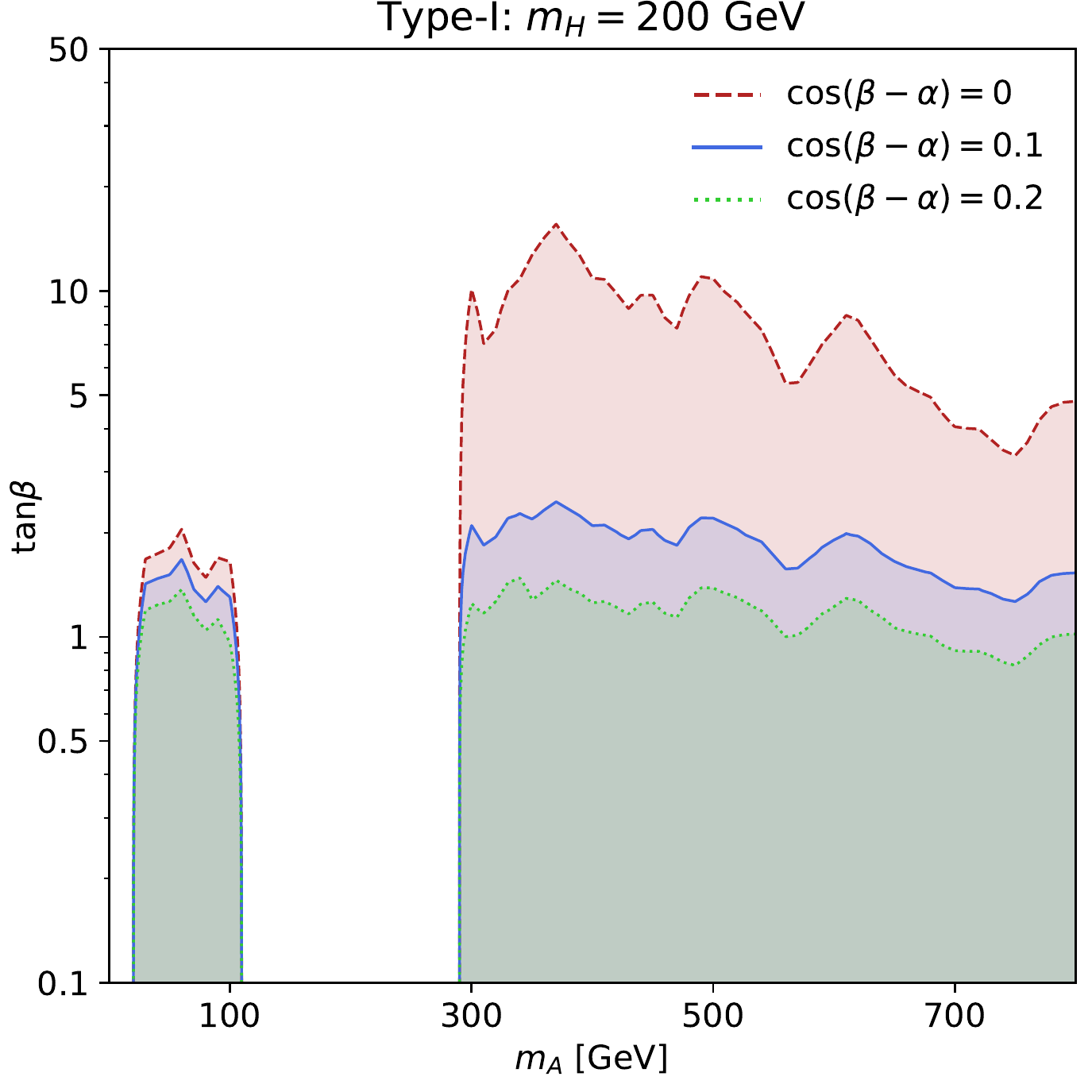}
\includegraphics[width=0.49\textwidth]{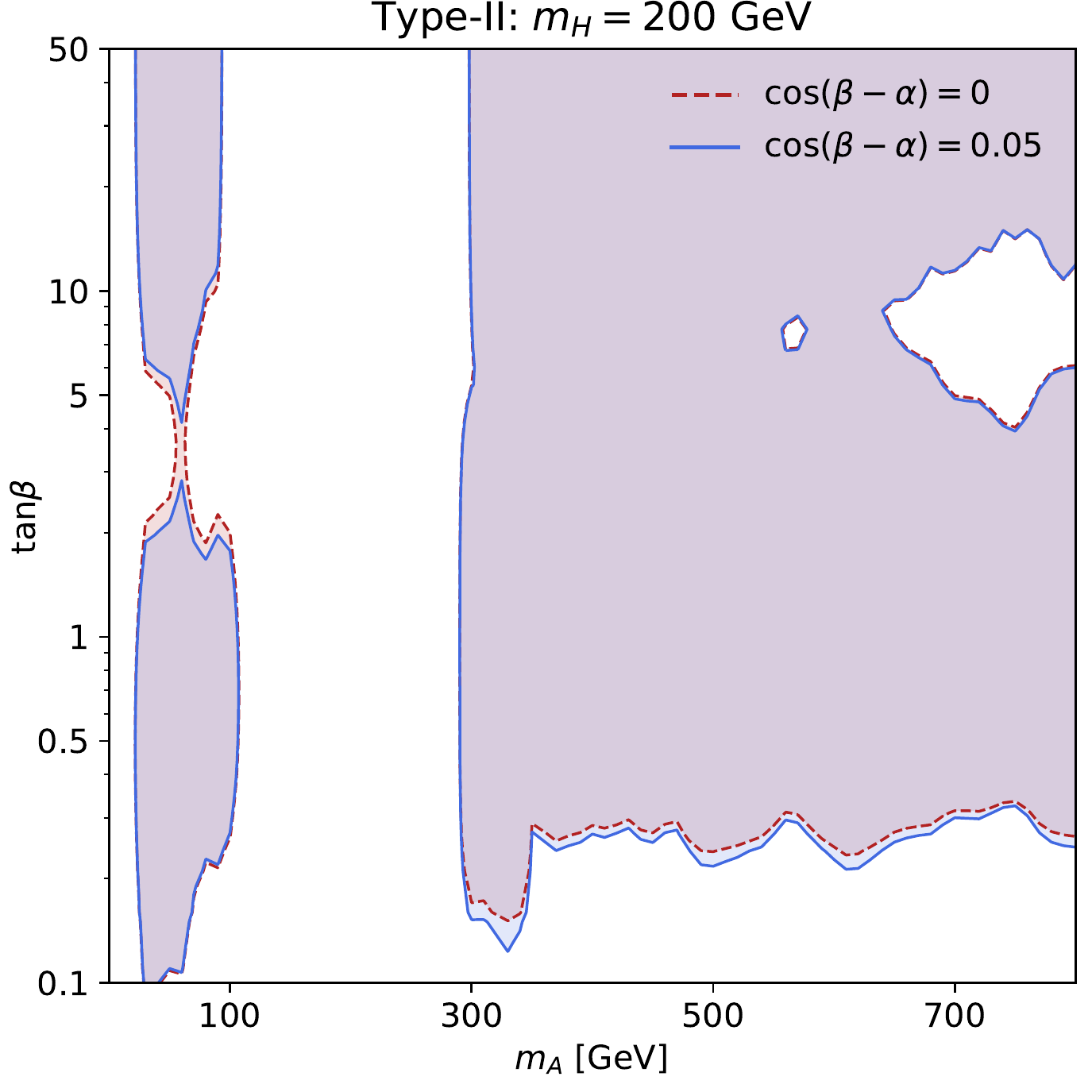}
\includegraphics[width=0.49\textwidth]{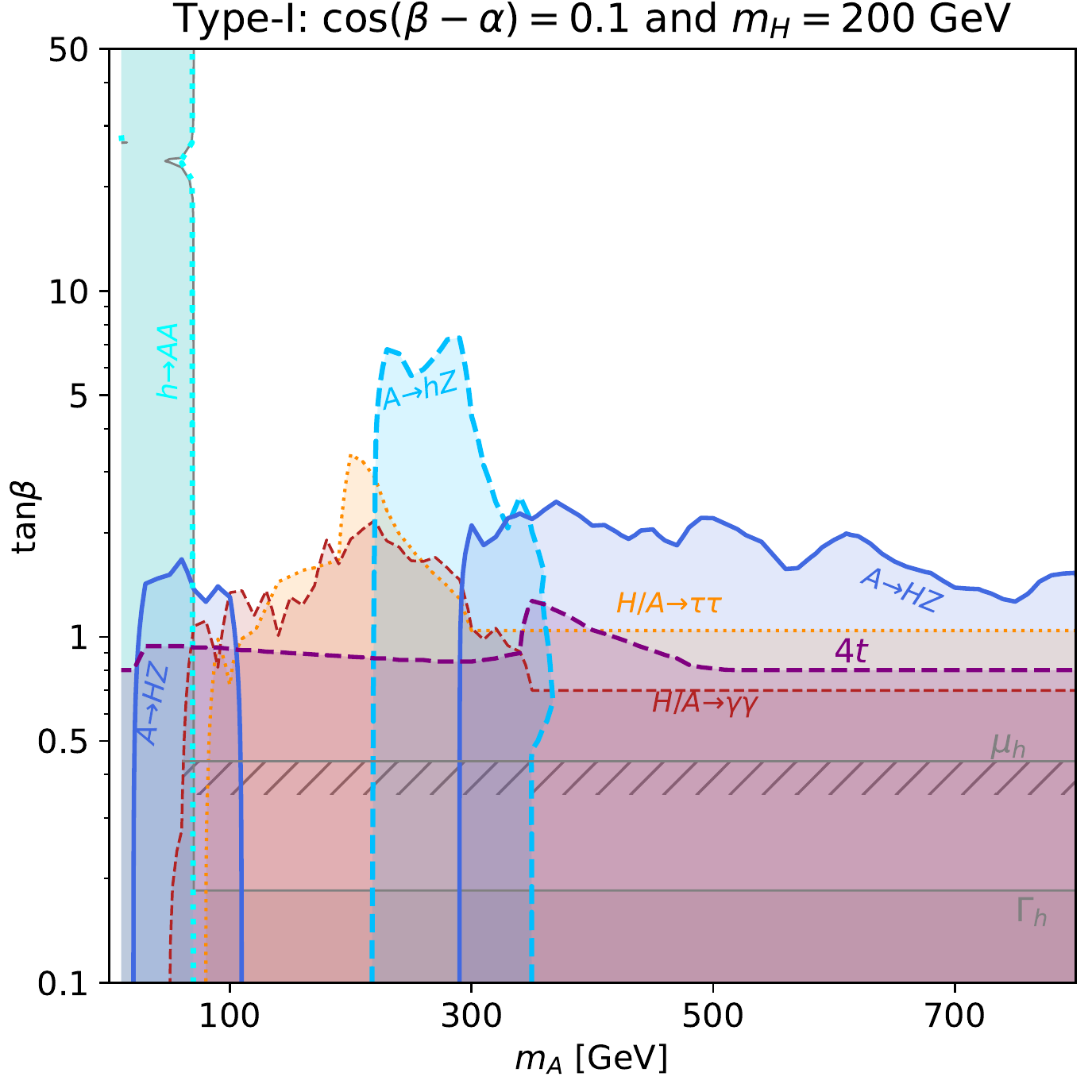}
\includegraphics[width=0.49\textwidth]{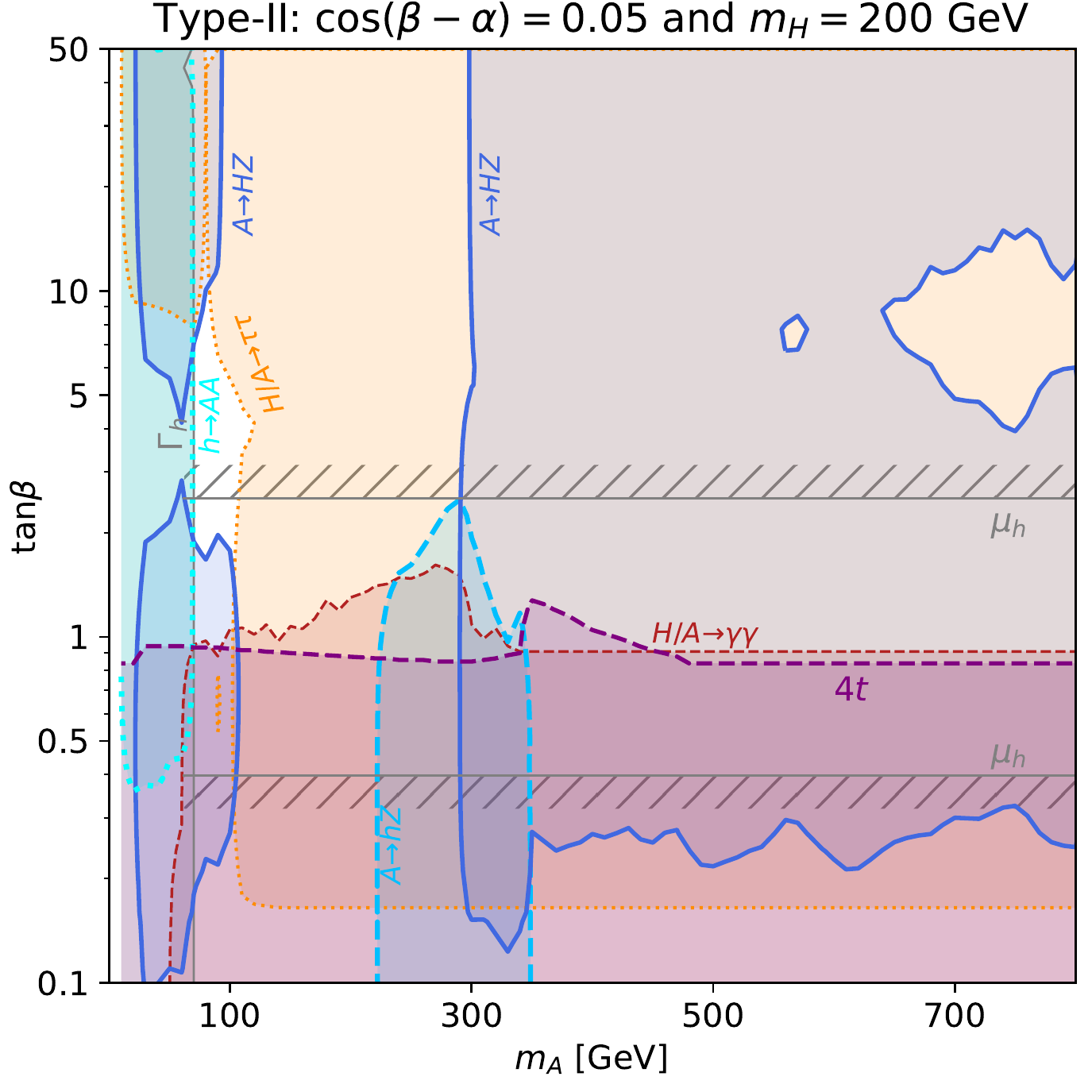}
\caption{
Constraints on the Type-I (left panel) and the Type-II (right panel) 2HDM in $m_A-\tb$ plane. 
\textbf{Top:} Parameter space excluded at 95\% C.L. by the $A/H \to HZ/AZ$ search for $m_H=200~\gev$ and $\cba=0$ (red), $0.1$ (blue) and $0.2$ (green) in the left panel and $\cba=0$ (red) and $0.05$ (blue) in the right panel.  
\textbf{Bottom:} Constraints at 95\% C.L. for $m_H=200~\gev$ and $\cba=0.1$ from LHC searches for $A \to HZ$ (blue), $A\to hZ$ (dashed dark blue), $H  \to AA$ (dashed cyan), $H/A \to \tau\tau$ (dotted orange), $H/A \to \gamma\gamma$ (dashed brown) and $4t$ production (dashed magenta) as well as the global fit of SM-like Higgs couplings strength $\mu_h$ (grey hatched region) and the Higgs width measurement $\Gamma_h \notin (0.08,9.16)~\mev$ (grey). 
}
\label{fig:matb}
\end{figure*}

While in \autoref{fig:conven}, we have considered the constraints in the $m_A$ vs. $\tb$ plane for a degenerate mass spectrum, we now consider the same parameter space again for a non-degenerate spectrum permitting the exotic decay channel $A/H\to HZ/AZ$. 

The top panel of \autoref{fig:matb} shows the constraints from the $A/H\to HZ/AZ$ channel for Type-I (left panel) and the Type-II (right panel) 2HDM for $m_H=200~\gev$. The low mass region $m_A < m_H - m_Z = 110~\gev$ probes the decay $H \to AZ$ while the high mass region  $m_A > m_H + m_Z = 290~\gev$ is sensitive to decay $A \to HZ$.

The largest reach is obtained under the alignment limit of $\cba=0$. For the Type-I 2HDM, $\tb$ up to about 10 can be excluded for $m_A > 290~\gev$, while the reach in $\tb$ is reduced when $\cba$ increases. For small $m_A$, when $H \to AZ$ is kinematically accessible, $\tb \lesssim 2$ can be excluded. Regions with large $\tb$ remain unconstrained, given the suppression of all the Yukawa couplings.   

For the Type-II 2HDM, the reach extends to large $\tb$ where the $bbA/H$ production rate is enhanced. Almost the entire range of $\tb$ is constrained by the $A/H\to HZ/AZ$ channel with the exceptions of the low $\tb<0.2$ region, where the branching fractions for $H/A \to bb$ and $\tau\tau$ are suppressed, and a gap at intermediate $\tb \sim 5$ at low and high masses $m_A$, where the Higgs production cross section is reduced. \medskip

The lower panels of \autoref{fig:matb} present the global constraints from direct search channels of the BSM Higgses, the Higgs coupling  $\mu_h$ and Higgs width $\Gamma_h$ precision measurements, and the $4t$ cross section measurements. While for $m_A \lesssim 300~\gev$, the strongest conventional search constrains are related to the decay of the pseudoscalar $A$, at large $m_A \gtrsim 300~\gev$ constraints mainly come from direct searches for $H$, whose mass is fixed to $m_H=200~\gev$. Therefore, there is no dependence on $m_A$ for the $\tau\tau$, $\gamma\gamma$ and 4$t$ exclusion limits in the large $m_A$ region.  
 
For the Type-I 2HDM with $\cba=0.1$, the small $m_A<m_h/2$ and the small $\tb \lesssim 2-3$ are excluded combining all channels, with the $A \to HZ$ gap of $|m_A-m_H|<m_Z$ region completely covered by the $A/H \to \tau\tau$ and $\gamma\gamma$, the $4t$, and the $A \to hZ$ channels. For the Type-II 2HDM with $\cba=0.05$, the small $\tb \lesssim 1$ region is covered mostly by the $H/A \to \gamma\gamma$, the $4t$, and the $A \to HZ$ and $hZ$ channels, while large $\tb$ region is covered by $H/A\to \tau\tau$ and $A\to HZ$. Note again that we have fixed $m_H=200~\gev$, which causes the $H\to \tau\tau$ channel to exclude the entire $\tb$ range for $m_A \gtrsim 100~\gev$.   Combinations of all channels exclude nearly the entire parameter space except for intermediate $\tb\sim 2$ with $m_A \sim 100~\gev$.

\section{Conclusion and Outlook}
\label{sec:con}

Since the discovery of 125 GeV Higgs boson, there have been many theoretical and experimental studies searching for additional scalar particles. In the framework of the 2HDM, besides the usual search methods for the non-SM Higgs bosons, such as SM Higgs precision measurements, conventional Higgs decays channels ($A/H \to f\bar f, VV, \gamma\gamma$), and decay to SM Higgs ($H\to hh, A \to hZ$), exotic decays of BSM Higgses such as $A/H \to HZ/AZ$ offer additional discovery potential when such modes are kinematically open. On the one hand, existing constraints based on the conventional searches will be relaxed given the reduced decay branching fractions.   On the other hand, such exotic decay modes provide additional search channels in the parameter regions with large mass splittings between non-SM Higgses.

In this study, we focused on the exotic decay of $A/H \to HZ/AZ$, which is the most promising channel given the large branching fraction as well as the clean experimental signal of $bb\ell\ell$ and $\tau\tau\ell\ell$.  ATLAS~\cite{Aaboud:2018eoy} and CMS~\cite{Sirunyan:2019wrn, Khachatryan:2016are} have performed searches for these decay channels at both 8 TeV and 13 TeV.  We performed a comprehensive study of the BSM neutral Higgses in the Type-I and Type-II 2HDMs under existing direct and indirect search results, and studied the constraints on their parameter spaces.  We list our main findings below.

\begin{itemize}
\item{}
Given the theoretical considerations and electroweak precision measurements, large mass splittings $\gtrsim 200$ GeV in BSM Higgs masses in the 2HDMs are generally not allowed for large BSM Higgs masses $\gtrsim 1$ TeV~\cite{Kling:2016opi}.  Therefore, the LHC is the most relevant machine to probe the parameter space of non-degenerate 2HDMs.  While the $bb\ell\ell$ final state is used in the current 13 TeV searches, final state with $\tau\tau\ell\ell$ will be promising at high luminosity given the reduced SM backgrounds comparing to the $bb\ell\ell$ mode.

\item{}
For the conventional search  channels, the most sensitive ones are $H/A \to \tau\tau$ and $\gamma\gamma$.  Other channels only provide subdominant constraints.  

\item{}
The low mass region of $m_{A/H}\sim 100~\gev$ is still challenging for both the conventional and exotic decay channels, motivating a continuation of searches in this mass region. For even lower masses of $m_{A/H}<m_h/2$, both the SM-like Higgs width measurements and $h \to AA/HH$ can be used to constrain this parameter region.  In particular, $h \to AA$ is unsuppressed even under the alignment limit. 

\item{}
Other than the resonant searches for the BSM Higgses, rate measurements of SM processes, for example, $4t$ or $ttZ$, can be used to constrain $ttH/ttA$  or $HZ/AZ$ production.  It is especially useful in the high mass region above the $t\bar{t}$ threshold when the decay widths of $H$ and $A$ are large and the resonant searches are ineffective. In particular, the usefulness of $ttZ$ channel motivates a dedicated search for $A \to HZ \to ttZ$ in the future.

\item{}
Combing all the searches, for the Type-I 2HDM, most of the $m_{A,H}<350$ GeV regions are excluded for low $\tanb$ under the alignment case.  Limits are reduced at intermediate and large $\tanb$ as well as $\cba\neq 0$.  For the Type-II 2HDM, the limits for large $\tanb$ are stronger, given the enhanced production cross sections. 

\item{}
The exotic decays modes, which have enhanced reach under the alignment limit, show great complementarity with the SM-like Higgs precision measurements, as well as the direct search modes of $VV$, $hZ$ and $hh$, which has reduced sensitivity near the alignment region. 

\item{}
The exotic decay mode of $A \to HZ$ extends the reach in $m_A$ and $\tanb$ beyond the conventional search channels.  Combining all the LHC search channels for the non-SM Higgses, for $m_H=200$ GeV, the entire region of $m_A$ up to about 800 GeV are excluded in the Type-I 2HDM for the low $\tanb$.   Almost the entire region of $m_A$ vs. $\tanb$ plane is excluded for the Type-II 2HDM except for a small region around  $\tanb\sim 2 $ and $m_A\sim 100~\gev$. \medskip
 
\end{itemize}

Besides minimal realizations of the 2HDM that we focused on in this paper, the exotic decay channel $A/H \to HZ/AZ$ channel has also been proven to be a useful probe for CPV realizations \cite{Hou:2019mve} and singlet extension \cite{Baum:2017gbj, Baum:2018zhf, Baum:2019uzg, Baum:2019pqc} of the 2HDM. Additionally, other exotic decay modes besides $A/H \to HZ/AZ$, such as $H^\pm \to AW, HW$, $A/H \to H^\pm W^\mp$, could also extend the reach beyond the direct search via conventional decay channels, and indirect reach via Higgs precision measurements, especially at 100 TeV $pp$ colliders~\cite{Kling:2018xud}.   Combing all the search channels will provide valuable information towards Higgs sectors beyond the SM.


\acknowledgments
WS is supported by the Australian Research Council Discovery Project DP180102209. FK is supported by the Department of Energy under Grant DE-AC02-76SF00515. SS is supported by the Department of Energy under Grant DE-SC0009913.  


\bibliographystyle{utphys}
\bibliography{references}

\end{document}